\documentclass[aps,prd,twocolumn,groupedaddress]{revtex4-2}

\usepackage[english]{babel}
\usepackage{graphicx}
\usepackage[table,dvipsnames]{xcolor}
\usepackage{float}
\usepackage{multirow}
\usepackage{makecell}
\usepackage{mathtools} 
\usepackage{physics} 

\allowdisplaybreaks

\newcommand{\RNumb}[1]{\textup{\uppercase\expandafter{\romannumeral#1}}}

\newcommand*{\linebreakvertS}[1]{%
\begingroup
   \begin{tabular}[c]{@{}>{\centering\arraybackslash}p{4.5cm}@{}}#1\end{tabular}%
  \endgroup
}

\newcommand*{\linebreakvertP}[1]{%
\begingroup
   \begin{tabular}[c]{@{}>{\centering\arraybackslash}p{2.5cm}@{}}#1\end{tabular}%
  \endgroup
}

\newcommand*{\linebreakvertD}[1]{%
\begingroup
   \begin{tabular}[c]{@{}>{\centering\arraybackslash}p{2.5cm}@{}}#1\end{tabular}%
  \endgroup
}

\begin{document}


\title{Relativistic description of asymmetric fully heavy tetraquarks in the diquark--antidiquark model}



\author{V. O. Galkin$^{1}$}
\email[]{galkin@ccas.ru}

\author{E. M. Savchenko$^{1,2}$}
\email[]{savchenko.em16@physics.msu.ru}

\affiliation{$^{1}$Federal Research Center ``Computer Science and Control'', Russian Academy of Sciences, Vavilov Street 40, 119333 Moscow, Russia \\ $^{2}$Faculty of Physics, M.V.Lomonosov Moscow State University, Leninskie Gory 1-2, 119991 Moscow, Russia}


\date{\today}

\begin{abstract} 
Masses of the ground, orbitally and radially excited states of the asymmetric fully heavy tetraquarks, composed of charm ($\rm c$) and bottom ($\rm b$) quarks and antiquarks are calculated in the relativistic diquark--antidiquark picture. The relativistic quark model based on the quasipotential approach and quantum chromodynamics is used to construct the quasipotentials of the quark--quark and diquark--antidiquark interactions. These quasipotentials consist of the short-range one-gluon exchange and long-distance linear confinement interactions. Relativistic effects are consistently taken into account. A tetraquark is considered as a bound state of a diquark and an antidiquark which are treated as a spatially extended colored objects and interact as a whole. It is shown that most of the investigated tetraquarks states (including all ground states) lie above the fall-apart strong decay thresholds into a meson pair. As a result they could be observed as wide resonances. Nevertheless, several orbitally excited states lie slightly above or even below these fall-apart thresholds, thus they could be narrow states.
\end{abstract}


\maketitle

\section{Introduction\label{Sec:Intro}}

\par
The possibility of the existence of hadrons with a content of valence quarks and antiquarks different from the quark--antiquark pair for mesons and three quarks for baryons has been considered since the first days of the existence of the quark model~\cite{GellMann1964, Zweig1964}. Nevertheless the lack of convincing experimental evidence of such multiquark states reduced interest in their study. However, in the last two decades the situation has completely changed, as the first explicit experimental evidence of such exotic hadrons has finally been obtained (see the extensive reviews~\cite{Review.2017, Review.2019, Review.2020} and references therein). Currently, a several dozens of exotic hadron candidates and a number of reliably confirmed tetraquarks $\rm qq\overline q\overline q$ ($\rm cs\overline u\overline d$ --- LHCb 2020~\cite{csud:LHCb:2020}; $\rm cu\overline d\overline s$, $\rm cd\overline u\overline s$ --- LHCb 2022~\cite{cuds;cdus:LHCb:2022}; $\rm cc\overline u\overline d$ --- LHCb 2021~\cite{ccud:LHCb:2021}; $\rm cu\overline c\overline s$ --- LHCb 2021~\cite{cucs;cscs:LHCb:2021}; $\rm cd\overline c\overline s$ --- LHCb 2023~\cite{cdcs:LHCb:2023}; $\rm cs\overline c\overline s$ --- CMS 2013~\cite{cscs:CMS:2013}, LHCb 2016, 2021, 2022~\cite{cscs:LHCb:2016, cucs;cscs:LHCb:2021, cscs:LHCb:2022}; $\rm cc\overline c\overline c$ --- LHCb 2020~\cite{cccc:LHCb:2020}, ATLAS 2023~\cite{cccc:ATLAS:2023}, CMS 2023~\cite{cccc:CMS:2023}) and pentaquarks $\rm qqqq\overline q$ ($\rm uudc\overline c$ --- LHCb 2015, 2019~\cite{uudcc:LHCb:2015, uudcc:LHCb:2019}; $\rm udsc\overline c$ --- LHCb 2022~\cite{udscc:LHCb:2022}) have been discovered. The most recent detailed review on the exotic hadrons can be found in Ref.~\cite{Review.2022}.
\par
Currently in the literature there is no consensus on the nature of experimentally observed states with exotic properties. Thus significantly different interpretations have been proposed for four-quark states. The main of them are the following.
\begin{itemize}

	\item Compact tetraquark (i.e. exotic hadron) consisting of a diquark and an antidiquark, bound by the strong color interactions~\cite{DiQ1993, TetraMaiani2005, Tetra2006};
	
	\item Molecule consisting of two mesons loosely coupled by the meson exchange~\cite{Mol2018};
	
	\item Hadroquarkonium consisting of a heavy quarkonium embedded in a light meson~\cite{HadroQ2008.1, HadroQ2008.2};
	
	\item Kinematic casp~\cite{Cusp2015}, etc.
	
\end{itemize}
\noindent
Discriminating among these description is a very difficult experimental task.
\par
The simplest exotic system is a tetraquark consisting of two quarks and two antiquarks. Heavy tetraquarks are of particular interest, because the presence of a heavy quark increases the binding energy of the bound system and, as a consequence, the probability that masses of such tetraquarks will be below the thresholds of decay into mesons with open or hidden heavy flavors. In such case the fall-apart strong decays occurring through the rearrangement of quarks and antiquarks are kinematically forbidden. Then the corresponding tetraquarks can decay only via the weak or electromagnetic interactions and, therefore, they must have a narrow decay width. If the predicted tetraquarks have masses slightly (by several MeV) exceeding these thresholds, then they could also be observed as resonances. The excited states of tetraquarks despite the large phase space could also be narrow resonances since their decays could be suppressed either by a centrifugal barrier between quarks and antiquarks for orbitally excited states or by nodes of the wave function for radially excited states or by both simultaneously. 
\par
\textcolor{black}{The fully heavy tetraquark states have been studied in the literature since the beginning of 80s (see e.g. Refs.~\cite{Chao1981, Ader1982, Heller1985, Zouzou1986, Badalian1987}).}
\par
In this paper we study the asymmetric in flavor fully heavy (containing only heavy quarks) tetraquarks. This choice significantly reduces the number of approaches used to describe them. There are already a number of theoretical calculations within the framework of a variety of models \textcolor{black}{\cite{qqQQ:udbb;ssbb.QQQQ:ccbb.2015.1, QQQQ.ground:all.2018.1, QQQQ.ground:all.2018.2, QQQQ.ground:cbcb;ccbb.2019.1, QQQQ:ccbb.2019.2, QQQQ.ground:all.2019.3, QQQQ.ground;exc:cccc;bbbb;ccbb.2019.4, qqQQ.ground:sbsb;ssbb;qqbb.QQQQ.ground;exc:cccc;cbcb;bbbb;ccbb.2020.1, QQQQ.ground:all.2020.2, QQQQ.ground:all.exc:ccbb.2021.1, QQQQ.ground:ccbb.2021.2, QQQQ.ground:all.2021.3, QQQQ.ground:all.2021.4, QQQQ.ground:cccc;cbcb;bbbb;ccbb.2022.1, QQQQ.ground:cccb;ccbb;bbbc.2022.2, QQQQ.ground:all.2022.3, QQQQ.ground:all.2023.1, qqqq;qqqQ;qqQQ;qQQQ;QQQQ.ground:all.1992.1, Richard2017, tetrons2018}}, but there is no consensus in them which of the predicted states can live long enough for their experimental detection.
\par
Experimental searches for the fully heavy tetraquarks are actively conducted at the Large Hadron Collider (LHC). And a certain successes have already been achieved. The LHCb~\cite{cccc:LHCb:2020}, ATLAS~\cite{cccc:ATLAS:2023} and CMS~\cite{cccc:CMS:2023} Collaborations have discovered the above mentioned fully charmed tetraquark $\rm cc\overline c\overline c$ (resonance X(6900), etc.), and its mass is consistent with our recent predictions from Refs.~\cite{Savch2020PhysRevD, Savch2021Uni, Savch2022Sym}. On the other hand, the searches for the fully bottom tetraquark $\rm bb\overline b\overline b$ by the LHCb~\cite{no.bbbb:LHCb:2018} and CMS~\cite{no.bbbb:CMS:2017, no.bbbb:CMS:2020} Collaborations has not yet yielded any results which is also consistent with our predictions. Note that production of the asymmetric fully heavy tetraquarks is much more difficult experimental task since it requires production of at least three heavy quark--antiquark pairs. Nevertheless, the search for the fully heavy tetraquarks of every possible flavor composition continues.
\par
The paper is organized as follows. In Sec.~\ref{Sec:Model} we give a description and physical justification of the model we have chosen for studying these four-quark structures. In Sec.~\ref{Sec:Math} we describe the Relativistic Quark Model and its application to the calculation of the tetraquark mass spectra. In Sec.~\ref{Sec:Res} we present the results of our calculations. In Sec.~\ref{Sec:Thr} we analyze our predictions by comparing them with the thresholds for the fall-apart strong decays into a pair of heavy mesons. In Sec.~\ref{Sec:Theor} we give a comparison of our results with the predictions of other researchers. Finally, in Sec.~\ref{Sec:Con} the results and conclusions are summarized.

\section{Model of Fully Heavy Tetraquarks\label{Sec:Model}}

\par
A tetraquark is a bound system of two quarks and two antiquarks $\rm q_{1}q_{2}\overline q_{3}\overline q_{4}$. There are 6 flavors of quarks which are classified by their current mass values~\cite{PDG2022} into two groups.
\begin{itemize}

	\item Light ($\rm u, d, s$) with $m_{\rm q} \ll \Lambda_{\rm QCD}$
	\begin{itemize}

		\item $m_{\rm u} = 2.16^{+0.49}_{-0.26}$ MeV, 
		
		\item $m_{\rm d} = 4.67^{+0.48}_{-0.17}$ MeV,
		
		\item $m_{\rm s} = 93.4^{+8.6}_{-3.4}$ MeV;
		
	\end{itemize}
	
	\item Heavy ($\rm c, b, t$) with $m_{\rm Q} \gg \Lambda_{\rm QCD}$
	\begin{itemize}

		\item $m_{\rm c} = 1.27\pm 0.02$ GeV, 
		
		\item $m_{\rm b} = 4.18^{+0.03}_{-0.02}$ GeV, 
		
		\item $m_{\rm t} = 172.69 \pm 0.30$ GeV.
	
	\end{itemize}
	
\end{itemize}
\noindent
where $\Lambda_{\rm QCD} \approx 200$ MeV is the quark confinement mass scale. In this paper we focus on the tetraquarks consisting only of heavy flavor quarks and antiquarks. However, we don't consider the $t$-quark as one of the possible constituents. It is almost two orders of magnitude heavier than the other heavy quarks, and, due to its colossal mass, it decays too quickly ($\tau_{\rm t} \lesssim 10^{-23}$ s) via the weak interaction not having enough time to form a bound state~\cite{Lifetime1986}.
\par
Two heavy quark flavors can be arranged into 9 types of fully heavy tetraquarks as follows.
\begin{itemize}

	\item 3 symmetric (with hidden charm and/or bottom)
	\begin{itemize}
	
		\item fully charmed $\rm cc \overline c\overline c$,
		
		\item doubly charmed--bottom $\rm cb\overline c\overline b$,
		
		\item fully bottom $\rm bb\overline b \overline b$;
		
	\end{itemize}
	
	\item 6 asymmetric (with open charm and bottom)
	
	\begin{itemize}
	
		\item triple charmed and bottom $\rm cc\overline c\overline b$, $\rm bc\overline c\overline c$,
		
		\item double charmed and double bottom $\rm cc \overline b\overline b$, $\rm bb\overline c\overline c$,

		\item triple bottom and charmed $\rm bb\overline b\overline c$, $\rm cb\overline b\overline b$.
		
	\end{itemize}
	
\end{itemize}
\noindent
Calculations for the ground states of all compositions~\cite{Savch2020PhysRevD, Savch2021Uni} and excited states of symmetric compositions~\cite{Savch2022Sym} have been done in previous papers. Here we continue our study and calculate excited states of asymmetric compositions.
\par
Tetraquarks are formed from the closely produced quark--antiquark pairs. Formation of symmetric combinations requires the production of only two pairs ($\rm 2 \times c\overline c$, $\rm c\overline c + b\overline b$ and $\rm 2 \times b\overline b$), while the formation of asymmetric combinations requires the production of at least three pairs which is a less probable event. Thus, symmetric states are more convenient for the experimental searches and we already have experimental candidates for the fully charmed tetraquarks (additional discussion on this topic can be found in Sec.~\ref{Sec:Thr}).
\par
We employ the diquark--antidiquark picture in which the tetraquark is considered as a bound state of two unobservable colored structures: the diquark $\rm [Q_{1}Q_{2}]$ and the antidiquark $\rm [\overline Q_{3} \overline Q_{4}]$. This model is not new and is widely used in the hadron spectroscopy giving good agreement between the calculations (for example, baryon masses) and experiments~\cite{Baryo2005, Regge2011baryo}.
\par
Another widely used approach for the description of four-quark states is the meson--meson molecular model. We consider such a picture to be significantly less probable for the fully heavy tetraquarks. Indeed, the binding in the molecule is mainly provided by the meson exchange resulting in the Yukawa-type potential. In the case of fully heavy tetraquarks such a meson contains only heavy quarks, thus, it is too massive to provide the sufficient binding (see discussion in Ref.~\cite{Savch2022Sym}, Sec.~2).
\par
When calculating in the diquark--antidiquark picture one must take into account that the diquark (the further discussion applies to the antidiquarks as well) is a bound system of fermions and therefore obeys the generalized Pauli principle requiring  the overall diquark wave function to be antisymmetric
\begin{equation}
\label{Eq:psid}
\Psi_{\rm diquark} = \psi_{\rm space} \times \psi_{\rm color} \times \psi_{\rm flavor} \times \psi_{\rm spin} \equiv \Psi_{\rm antisym.}.
\end{equation}
\begin{itemize}

	\item The parity of the spatial part of the wave function is determined by the angular momentum $L$. Considering only ground states of the diquarks (the justification of this assumption will be given further in Sec.~\ref{Sec:Res}) and allowing excitations in the diquark--antidiquark system we get
	\begin{equation}
	\label{Eq:psidspace}
	\begin{gathered}
	P_{\rm ground \ state} = (-1)^{L_{\rm ground \ state}} \equiv (-1)^{0} = 1,
	\\
	\Longrightarrow \psi_{\rm space} \equiv \psi_{\rm sym.}.
	\end{gathered}
	\end{equation}
	
	\item The color part of the wave function depends on the choice of the color representation of the diquark. Two quarks are two color triplets, and in combination they can give either a color symmetric sextet or an antisymmetric antitriplet (symmetrically-reversed for antidiquarks)
	\begin{equation}
	\label{Eq:qqcolor}
	\begin{gathered}
	3 \times 3 = 6 \oplus \overline{3},
	\\
	\overline{3} \times \overline{3} = \overline{6} \oplus 3.
	\end{gathered}
	\end{equation}
	\noindent
	Note that in the color-sextet representation the internal interaction between the quarks within the diquark is repulsive, since the mean value of the product of the color SU(3) generators between sextet states is positive. Therefore, the color-sextet diquark cannot be a bound state. Contrarily, the interaction between the quarks within the color-antitriplet diquark is attractive, thus, it is reasonable to choose the color-antitriplet for the further considerations. As a result
	\begin{equation}
	\label{Eq:psidcolor}
	\psi_{\rm color} \equiv \psi_{\rm antisym.}.
	\end{equation}
	\textcolor{black}{Note that if the quark--antiquark cross diquark--antidiquark interactions are included then they could stabilize the sextet--antisextet diquarks in the tetraquark. However, such interactions are neglected in the pure diquark--antidiquark model which we adopt here. This approximation reduces the Hilbert space leading to an upper limit of the full result.}
	
	\item Thus, the remaining combination of the flavor and spin parts of the diquark wave function should be symmetric
	\begin{equation}
	\label{Eq:psidspinflavor} 
	\psi_{\rm flavor} \times \psi_{\rm spin} \equiv \psi_{\rm sym.}.
	\end{equation}
	\noindent
	Therefore, two combinations are allowed
	\begin{equation}
	\label{Eq:psidspinpsidflavor} 
	\begin{cases}
		\psi_{\rm flavor} = \psi_{\rm sym.},
		\\ 
		\psi_{\rm spin} = \psi_{\rm sym.}, 
	\end{cases} \quad {\rm or} \quad
	\begin{cases}
		\psi_{\rm flavor} = \psi_{\rm antisym.},
		\\ 
		\psi_{\rm spin} = \psi_{\rm antisym.}.
	\end{cases}
	\end{equation}
	
\end{itemize}
\noindent
This means that if a diquark consists of quarks of the same flavor with the symmetric flavor part it can only be axialvector with the symmetric spin part. If a diquark contains quarks of different flavors then both symmetric and antisymmetric flavor parts are possible and the diquark can be both axialvector (A) and scalar (S) with  spins $S_{\rm d}=1$ and $S_{\rm d}=0$, respectively.

\section{Relativistic Diquark--Antidiquark Picture\label{Sec:Math}}

\par
The considered tetraquarks should be treated relitivistically. Indeed, the estimates from Ref.~\cite{Mes2003} show that heavy quark velocities can reach up to half the speed of light. In particular, we use the relativistic kinematics and dynamics provided by the Relativistic Quark Model based on the quasipotential approach which has successfully been applied for calculating the mass spectra of ordinary three-quark baryons~\cite{Regge2011baryo} and quark--antiquark mesons~\cite{Regge2011meson}. Moreover, for the calculation of the tetraquarks mass spectra we use the diquark--antidiquark picture of tetraquarks and treat constituent diquarks as non-pointlike, spatially extended objects which interact as a whole. Thus, we account for the diquark internal structure and do not consider cross interactions between a quark from the diquark with an antiquark from the antidiquark. 
\par
In this framework the mass of a bound state is the solution of the relativistic Schr\"odinger-type quasipotential equation~\cite{Logunov1963, Faustov1985eng, Faustov1990eng} which describes this bound state of two particles in a given quasipotential. In particular, we reduce the four-body problem to two consecutive two-body problems, as we first apply this approach to the quark--quark system forming a diquark~\cite{Baryo2002, Baryo2005, Tetra2006, Tetra2007}, and then to the diquark--antidiquark system forming a tetraquark~\cite{Tetra2006, Tetra2007}. The quasipotential equation has the following form
\begin{widetext}
\begin{equation}
\label{Eq:Schr}
\bigg( \frac{b^{2}(M)}{2\mu_{\rm R}(M)}-\frac{\mathbf{p}^{2}}{2\mu_{\rm R}(M) } \bigg) \Psi_{\rm d,T}(\mathbf{p}) = \int \frac{d^{3}q}{(2\pi)^{3}}\: V(\mathbf{p}, \mathbf{q}; M) \Psi_{\rm d,T}(\mathbf{q}).
\end{equation}
\end{widetext}
\noindent
Here:
\begin{itemize}

	\item indexes $\rm d$ and $\rm T$ denote the diquark and the tetraquark, respectively; 
	
	\item $\mathbf{p}$ is a vector of the relative momentum;
	
	\item $M$ is the mass of the bound state;
	
	\item $\mu_{\rm R}$ is the relativistic reduced mass of the constituents, given by
	\begin{equation}
	\label{Eq:muR}
	\mu_{\rm R} = \frac{E_{1}E_{2}}{E_{1}+E_{2}} = \frac{M^4-(m_{1}^{2} - m_{2}^{2})^{2}}{4M^{3}},
	\end{equation}
	\noindent
	where:
	\begin{itemize}
	
		\item $m_{1, 2}$ are masses of the constituents,
		
		\item $E_{1, 2}$ are the on-mass-shell energies of the constituents
		\begin{equation}
		\label{Eq:En}
		E_{1, 2} = \frac{M^{2}+m_{1, 2}^{2}-m_{2, 1}^{2}}{2M};
		\end{equation}
		
	\end{itemize}
	
	\item $b^{2}(M)$ is the on-mass-shell relative momentum in the center-of-mass system squared
	\begin{equation}
	\label{Eq:b^2}
	b^{2}(M)  = \frac{[M^{2}-(m_{1}+m_{2})^{2}][M^{2}-(m_{1}-m_{2})^{2}] }{4M^{2}};
	\end{equation}
	
	\item $\Psi_{\rm T, d}(\mathbf{p})$ are the bound state wave functions;
	
	\item $V(\mathbf{p}, \mathbf{q}; M)$ is the quasipotential operator of the constituents.
	
\end{itemize}
\par
Indeed, Eq.~\eqref{Eq:Schr} is relativistic. Its left-hand side contains relativistic kinematics, since the reduced mass of the bound state $\mu_{\rm R}$ and the on-mass-shell relative momentum $b^{2}(M)$ are complicated functions of the bound state mass $M$ (Eqs.~\eqref{Eq:muR},~\eqref{Eq:b^2}). The relativistic dynamics is contained in the right-hand side of Eq.~\eqref{Eq:Schr}, as the quasipotential $V(\mathbf{p}, \mathbf{q}; M)$ is constructed with the help of the off-mass-shell scattering amplitude, projected onto the positive-energy states and contains all relativistic spin-independent and spin-dependent contributions.
\par
The quasipotential of the quark-quark interaction in a diquark, as a constituent of the tetraquark, was already thoroughly discussed in, e.g., Ref.~\cite{Savch2022Sym} (see Eqs.~(5)-(15)). On its basis the diquark masses were already calculated in Refs.~\cite{Baryo2002, Baryo2005, Tetra2006} which we use here as an input. Thus we do not give here further details. On the other hand, calculation of the tetraquark mass spectra is the main goal of this paper. Therefore, we discuss the construction of the diquark--antidiquark quasipotential in detail. First, we use the same assumptions about the structure of the interaction as for the quark--quark case. The effective interaction is the sum of the usual one-gluon exchange term with the mixture of the long-range vector and scalar linear confining potentials, where the vector confining potential vertex contains the additional Pauli term. Second, we take into account the finite size of the diquarks and their integer spin. As a result, the quasipotential is given by~\cite{Tetra2006, Tetra2007}
\begin{widetext}
\begin{align}
V(\mathbf{p}, \mathbf{q}; M) & = \underbrace{\frac{\bra{d(\mathcal{P})} J_{\mu} \ket{d(\mathcal{Q})}}{2\sqrt{E_{\rm d}}\sqrt{E_{\rm d}}} \frac{4}{3}\alpha_{\rm s}D^{\mu\nu}(\mathbf{k}) \frac {\bra{d'(\mathcal{P'})} J_{\nu} \ket{d'(\mathcal{Q'})}}{2\sqrt{E_{\rm d'}}\sqrt{E_{\rm d '}}}}_{\substack{\rm diquark-gluon \; interation}} \nonumber
\\[5pt]
& + \underbrace{\Psi_{\rm d}^{*}(\mathcal{P})\Psi_{\rm d'}^{*}(\mathcal{P'}) [J_{\rm d; \mu}J_{\rm d'}^{\mu}V_{\rm conf.}^{\rm V}(\mathbf{k})+V_{\rm conf.}^{\rm S}(\mathbf{k})]\Psi_{\rm d}(\mathcal{Q})\Psi_{\rm d'}(\mathcal{Q}')}_{\substack{\rm confinement}}, \label{Eq:Vdd}
\end{align}
\end{widetext}
\noindent
where the first term dominates at short distances while the second one is dominant at long distances. The main ingredients of Eq.~\eqref{Eq:Vdd} are as follows (for more details and their explicit forms see Eqs.~(10)-(12), (16)-(23) in Ref.~\cite{Savch2022Sym}).
\begin{itemize}

	\item $\rm d$ and $\rm d'$ denote the diquark and the antidiquark (i.e. constituents);
	
	\item ${\mathcal{Q}=(E_{\rm d}, \; \pm\mathbf{q})}$ and ${\mathcal{P}=(E_{\rm d}, \; \pm \mathbf{p})}$ are the initial and final diquark momenta, respectively;
	
	\item $k=\mathcal{P}-\mathcal{Q}=(0, \mathbf{k})$;
	
	\item $E_{\rm d, d'}$ are the on-shell diquark energies defined by Eq.~\eqref{Eq:En}, where $m_{1,2} \equiv M_{\rm d,d'}$ are the diquark and antidiquark masses and $M$ is the tetraquark mass;
	
	\item $\alpha_{\rm s}$ is the running QCD coupling constant with freezing~\cite{alphasSimonov, alphasBadalian} that takes into account the number of open quark flavors and energy scale as the reduced constituents mass;
	
	\item $D_{\mu\nu}(\mathbf{k})$ is the gluon propagator in the Coulomb gauge;
	
	\item $\Psi_{\rm d}(\mathcal{P})$ is the diquark wave function which is unit matrix for the scalar diquark and polarization vector for the axialvector diquark;
	
	\item $J_{\rm d; \mu}$ is the effective long-range vector interaction vertex of the diquark with the total chromomagnetic moment of the diquark chosen to be equal to zero to vanish the long-range chromomagnetic interaction. 

\end{itemize}
\noindent
Let us briefly discuss the most important terms.
\begin{itemize}

	\item $V_{\rm conf.}^{\rm V,S}$ are the vector and scalar confining potentials which in the nonrelativistic limit in configuration space have the form consistent with the lattice calculations
	\begin{align}
	& V_{\rm conf.}^{\rm V}(r) = (1-\varepsilon)V_{\rm conf.}(r), \nonumber
	\\[3pt]
	& V_{\rm conf.}^{\rm S}(r) = \varepsilon V_{\rm conf.}(r), \label{Eq:conf}
	\\[3pt]
	& V_{\rm conf.}(r) = V_{\rm conf.}^{\rm V}(r)+V_{\rm conf.}^{\rm S}(r) = Ar+B, \nonumber
	\end{align}
	\noindent
	where $\varepsilon$ is the mixing coefficient. 
	
	\item $\bra{d(\mathcal{P})} J_{\mu} \ket{d(\mathcal{Q})}$ is the diquark--gluon interaction vertex that accounts for the internal structure of the diquark and leads to the emergence of the form factor $F(r)$ smearing the one-gluon exchange potential (see Fig.~1 and Eqs.~(24)-(27) in Ref.~\cite{Savch2022Sym}).
	
\end{itemize}
\par
It is necessary to calculate the matrix elements of quark currents between diquarks $\bra{d(\mathcal{P})} J_{\mu} \ket{d(\mathcal{Q})}$ to take into account the finite size of the diquark. These matrix elements are elastic (diagonal) and can be parametrized by the set of form factors $h_{+,1,2,3}(k^{2})$ (see Eqs.~(28)-(29) in Ref.~\cite{Savch2022Sym}). They, in turn, can be expressed by the form factor in the momentum space $F(\mathbf{k}^{2})$ calculated as the overlap integral of the diquark wave functions (see Eqs.~(30)-(31) in Ref.~\cite{Savch2022Sym}). \color{black}The form factor $F(r)$, which accounts for the diquark size, was first introduced for the description of doubly heavy baryons in Ref.~\cite{Baryo2002} as the Fourier transform of the $\frac{F(\mathbf{k}^{2})}{\mathbf{k}^{2}}$ multiplied then by $r$. It can be parametrized with high accuracy as~\cite{Baryo2002}
\begin{equation}
\label{Eq:Fr}
F(r) = 1 - e^{-\xi r - \zeta r^{2}}.
\end{equation} \color{black} 
\begin{figure}
\includegraphics[width=9.1cm]{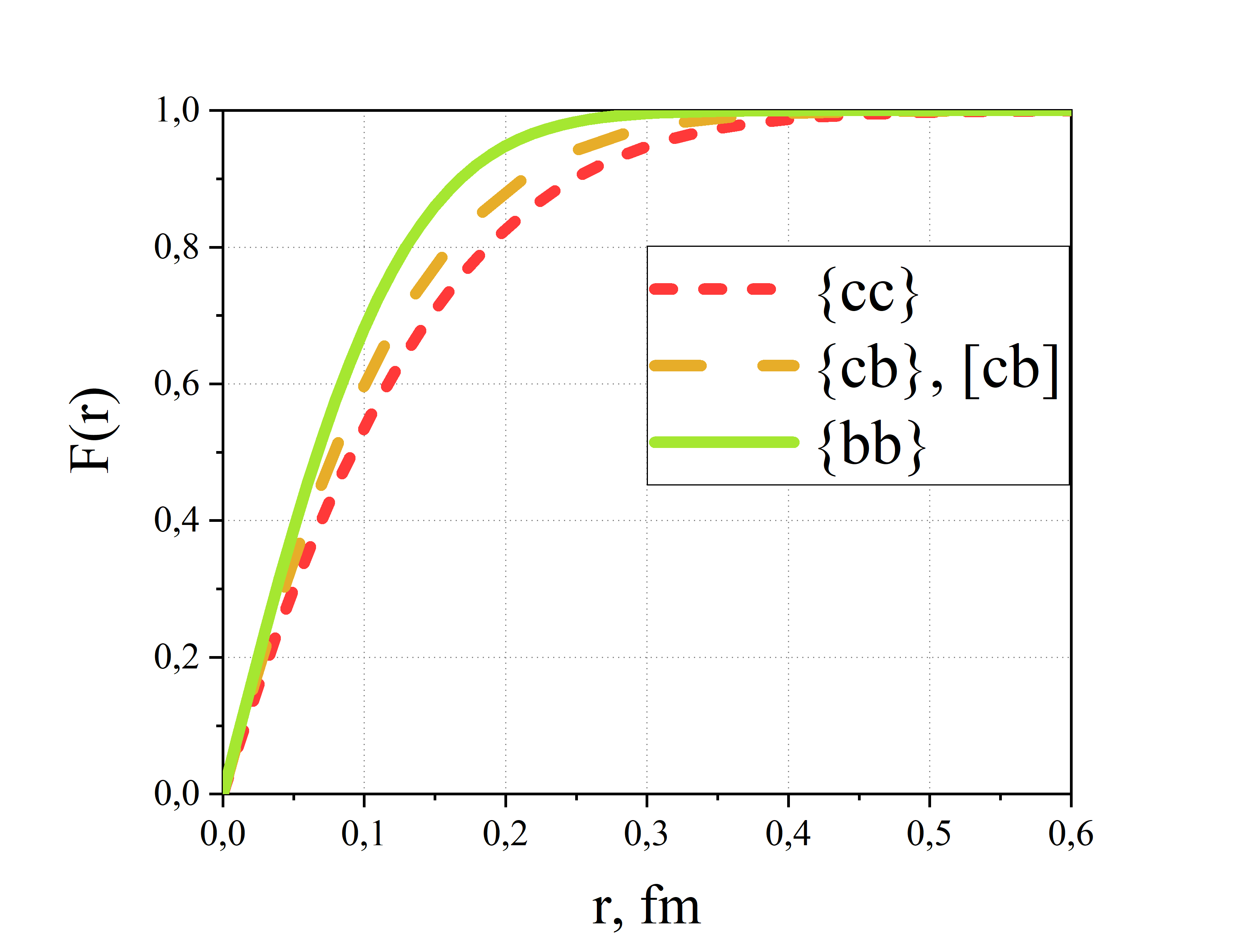}
\caption{Form factors $F(r)$ for the various doubly heavy diquarks. $\rm \{Q,Q' \}$ denotes the axialvector and $\rm [Q,Q']$ denotes the scalar diquarks, respectively.\label{Fig:FormF_QQQQ}}
\end{figure}
\color{black} 
These form factors for the doubly heavy diquarks are shown in Fig.~\ref{Fig:FormF_QQQQ}. In the tetraquark case the convolution of the diquark and antidiquark form factors arises. The numerical calculation shows that it can be well approximated by modifying the Coulomb potential via the diquark and antidiquark form factors in the following way
\begin{equation} 
\label{Eq:Coul}
V_{\rm Coul.}(r) \equiv -\frac{4}{3}\alpha_{\rm s}\frac{F_{1}(r)F_{2}(r)} {r}.
\end{equation}
\noindent
As can be seen from Fig.~\ref{Fig:FormF_QQQQ} the closer we are to the doubly heavy diquark the more of its internal structure is resolved, while at large distances ($r>0.5$ fm) its internal structure cannot be resolved and it looks point-like. Thus it is not necessary to introduce the diquark form factor into the confining interaction.  \color{black} 
\par
Finally, we obtain the diquark--antidiquark interaction potential~\cite{Tetra2007, Savch2021Uni, Savch2022Sym}
\vspace*{-3.0\baselineskip}
\stepcounter{equation}
\begin{widetext}
\begin{align}
V(r) = \textcolor{LimeGreen}{\pmb{\Bigg[}} &V_{\rm Coul.}(r)+V_{\rm conf.}(r) + \frac{1}{E_{1}E_{2}} \Bigg\{ \mathbf{p}\bigg[ V_{\rm Coul.}(r)+V_{\rm conf.}^{\rm V}(r)\bigg] \mathbf{p} -\frac{1}{4}\Delta V_{\rm conf.}^{\rm V}(r) + V_{\rm Coul.}^{'}(r)\frac{\mathbf{L}^{2}}{2r} \Bigg\} \textcolor{LimeGreen}{\pmb{\Bigg]}} \label{Eq:Va} \tag{{\theequation}.1}
\\[1pt]
+ \textcolor{LimeGreen}{\pmb{\Bigg[}} &\Bigg\{ \frac{1}{2} \bigg[ \frac{1}{E_{1}(E_{1}+M_{1})} + \frac{1}{E_{2}(E_{2}+M_{2})} \bigg] \frac{V_{\rm Coul.}^{'}(r)}{r}- \frac{1}{2} \bigg[ \frac{1}{M_{1}(E_{1}+M_{1})} + \frac{1}{M_{2}(E_{2}+M_{2})} \bigg] \frac{V_{\rm conf.}^{'}(r)}{r}  \nonumber
\\[1pt]
& \hspace{-1.5mm}+ \frac{\mu_{\rm d}}{4} \bigg[ \frac{1}{M_{1}^{2}} + \frac{1}{M_{2}^{2}} \bigg] \frac{V_{\rm conf.}^{'V}(r)}{r} + \frac{1}{E_{1}E_{2}} \bigg[ V_{\rm Coul.}^{'}(r) + \frac{\mu_{\rm d}}{4}\Big( \frac{E_{1}}{M_{1}}+\frac{E_{2}}{M_{2}}\Big) V_{\rm conf.}^{'V}(r) \bigg] \frac{1}{r} \Bigg\} \mathbf{L}(\mathbf{S_{1}}+\mathbf{S_{2}}) \nonumber 
\\[1pt]
+ &\Bigg\{ \frac{1}{2} \bigg[ \frac{1}{E_{1}(E_{1}+M_{1})} - \frac{1}{E_{2}(E_{2}+M_{2})} \bigg] \frac{V_{\rm Coul.}^{'}(r)}{r} - \frac{1}{2} \bigg[ \frac{1}{M_{1}(E_{1}+M_{1})} - \frac{1}{M_{2}(E_{2}+M_{2})} \bigg] \frac{V_{\rm conf.}^{'}(r)}{r} \nonumber
\\[1pt]
& \hspace{-1.5mm}+ \frac{\mu_{\rm d}}{4} \bigg[ \frac{1}{M_{1}^{2}} - \frac{1}{M_{2}^{2}} \bigg] \frac{V_{\rm conf.}^{'V}(r)}{r} + \frac{1}{E_{1}E_{2}} \frac{\mu_{\rm d}}{4}\Big( \frac{E_{1}}{M_{1}}-\frac{E_{2}}{M_{2}}\Big) \frac{V_{\rm conf.}^{'V}(r)}{r} \Bigg\} \mathbf{L}(\mathbf{S_{1}}-\mathbf{S_{2}}) \textcolor{LimeGreen}{\pmb{\Bigg]}} \label{Eq:Vb} \tag{{\theequation}.2}
\\[1pt]
+ \textcolor{LimeGreen}{\pmb{\Bigg[}} &\frac{1}{3E_{1}E_{2}} \Bigg\{ \frac{1}{r}V_{\rm Coul.}^{'}(r)-V_{\rm Coul.}^{''}(r) + \frac{\mu_{\rm d}^{2}}{4}\frac{E_{1}E_{2}}{M_{1}M_{2}}\Big( \frac{1}{r}V_{\rm conf.}^{'V}(r)-V_{\rm conf.}^{''V}(r)\Big) \Bigg\} \bigg[ \frac{3}{r^{2}}\Big( \mathbf{S_{1}r}\Big) \Big(\mathbf{S_{2}r}\Big) -\mathbf{S_{1}S_{2}}\bigg] \textcolor{LimeGreen}{\pmb{\Bigg]}} \label{Eq:Vc} \tag{{\theequation}.3}
\\[1pt]
+ \textcolor{LimeGreen}{\pmb{\Bigg[}} &\frac{2}{3E_{1}E_{2}} \Bigg\{ \Delta V_{\rm Coul.}(r) + \frac{\mu_{\rm d}^{2}}{4} \frac{E_{1}E_{2}}{M_{1}M_{2}} \Delta V_{\rm conf.}^{\rm V}(r) \Bigg\} \mathbf{S_{1}S_{2}} \textcolor{LimeGreen}{\pmb{\Bigg]}}, \label{Eq:Vd} \tag{{\theequation}.4}
\\
\label{Eq:V} \tag{\theequation}
\end{align}
\end{widetext}
\noindent
where indexes $1,2$ denote the diquark and antidiquark, $\mathbf{p}$ is the relative momentum, $M_{1, 2}$ and $E_{1, 2}$ are the masses and energies of the diquark and antidiquark, $\mu_{\rm d}$ is the total chromomagnetic moment of the diquark (we chose it to be zero), $\mathbf{S_{1,2}}$ is the diquark spin, $\mathbf{L}$ is the relative orbital momentum of the system, $V_{\rm conf.}$ is the confining potential in the nonrelativistic limit~\eqref{Eq:conf}. 
\par
The quasipotential~\eqref{Eq:V} contains the spin-independent~\eqref{Eq:Va} including orbit--orbit interaction and spin-dependent terms, namely, spin--orbit interaction~\eqref{Eq:Vb}, tensor interaction~\eqref{Eq:Vc} and spin--spin interaction \eqref{Eq:Vd}. To evaluate the mass of the state with particular quantum numbers we need to calculate the spin-orbit matrix elements of the quasipotential. Employing the known relations between the orbital momentum $\mathbf{L}$, total $\mathbf{S}$ and constituent spins $\mathbf{S_{1,2}}$ and the total momentum $\mathbf{J}$
\begin{equation}
\begin{gathered}
\label{Eq:Mom}
\mathbf{J} = \mathbf{L} + \mathbf{S}, 
\\
\mathbf{S} = \mathbf{S_{1}} + \mathbf{S_{2}},
\end{gathered}
\end{equation}
\noindent
and known matrix elements of the momenta squared
\begin{equation}
\begin{gathered}
\label{Eq:Matr}
\bra{\rm L \ S \ J} \mathbf{K^{2}} \ket{\rm L' \ S' \ J'} =  K(K+1) \cdot \delta_{L,L'} \delta_{S,S'} \delta_{J,J'}, 
\\
\mathbf{K} = \mathbf{L}, \mathbf{S}, \mathbf{S_{1,2}}, \mathbf{J},
\end{gathered}
\end{equation}
\noindent
we obtain the following expressions for the considered spin--orbit matrix elements (for details see Ref.~\cite{DeShalit2004}).
\begin{itemize}

	\item Orbit--orbit interaction $\bigg[ \mathbf{L}^{2} \bigg] \equiv \rm LL$:
	\begin{equation}
	\label{Eq:LL}
	\bra{\rm L \ S \ J} \mbox{\rm LL} \ket{\rm L' \ S' \ J'} = L(L+1) \cdot \delta_{L,L'} \delta_{S,S'} \delta_{J,J'}.
	\end{equation}
	
	\item Spin--orbit interactions $\bigg[ \mathbf{L}(\mathbf{S_{1}} \pm \mathbf{S_{2}}) \bigg] \equiv \rm LS_{\pm}$:
	\begin{widetext}
	\begin{align}
	\bra{\rm L \ S \ J} {\rm LS_{\pm}} \ket{\rm L' \ S' \ J'} = & (-1)^{L+S+J+S_{1}+S_{2}+1} \sqrt{L(L+1)(2L+1)(2S+1)(2S'+1)}
		\begin{Bmatrix}
		L & S & J\\
		S' & L & 1
		\end{Bmatrix} \nonumber 
	\\
	\times \Bigg( & (-1)^{S'} \sqrt{S_{1}(S_{1}+1)(2S_{1}+1)} 
		\begin{Bmatrix}
		S_{1} & S_{1} & 1\\
		S' & S & S_{2}
		\end{Bmatrix} \nonumber
	\\
	\pm & (-1)^{S} \sqrt{S_{2}(S_{2}+1)(2S_{2}+1)} 
		\begin{Bmatrix}
		S_{2} & S_{2} & 1\\
		S' & S & S_{1}
		\end{Bmatrix} \Bigg) \cdot \delta_{L,L'} \delta_{J,J'}. \label{Eq:LS}
	\end{align}
	\end{widetext}
	
	\item Tensor interaction $\bigg[ \dfrac{3}{r^{2}}\Big( \mathbf{S_{1}r}\Big) \Big(\mathbf{S_{2}r}\Big) -\mathbf{S_{1}S_{2}} \bigg] \equiv \rm T$:
	\begin{widetext}
	\begin{align}
	\bra{\rm L \ S \ J} {\rm T} \ket{\rm L' \ S' \ J'} = (-1)^{S+J} & \sqrt{30(2L+1)(2L'+1)(2S+1)(2S'+1)} \nonumber
	\\
	\times & \sqrt{S_{1}(S_{1}+1)(2S_{1}+1)} \sqrt{S_{2}(S_{2}+1)(2S_{2}+1)} \nonumber
	\\	
		\times & \begin{pmatrix}
		L' & 2 & L\\
		0 & 0 & 0
		\end{pmatrix}
		\begin{Bmatrix}
		J & S' & L'\\
		2 & L & S
		\end{Bmatrix}
		\begin{Bmatrix}
		S_{1} & S_{2} & S\\
		S_{1} & S_{2} & S'\\
		1 & 1 & 2
		\end{Bmatrix} \cdot \delta_{J,J'}. \label{Eq:T}
	\end{align}
	\end{widetext}
	 
	\item Spin--spin interaction $\bigg[ \mathbf{S_{1}S_{2}} \bigg] \equiv \rm SS$:
	\begin{widetext}
	\begin{equation}
	\label{Eq:SS}
	\bra{\rm L \ S \ J} \mbox{\rm SS} \ket{\rm L' \ S' \ J'} = \dfrac{1}{2} \Big( S(S+1) - S_{1}(S_{1}+1) - S_{2}(S_{2}+1) \Big) \cdot \delta_{L,L'} \delta_{S,S'} \delta_{J,J'}.
	\end{equation}
	\end{widetext}
	
\end{itemize}
\noindent
Eqs.~\eqref{Eq:LS}-\eqref{Eq:T} contain 3j-, 6j- and 9j-symbols which can be found in the literature~\cite{DeShalit2004}.
\par
Note, that the calculation of the masses of tetraquarks with the asymmetric compositions requires additional account for the significant mixing between the states with the same observable quantum numbers (total momentum and parity) $J^{P}$, but different full tetraquark spins $S,S'$ within the same excitation $n\rm L$: $n \prescript{2S+1}{}{\rm L}_{J} \leftrightarrow n \prescript{2S'+1}{}{\rm L}_{J}$. In particular, the spin--orbit matrix elements for the orbit--orbit $\rm LL$-~\eqref{Eq:LL}, spin--orbit $\rm LS_{+}$-~\eqref{Eq:LS} and spin--spin $\rm SS$-~\eqref{Eq:SS} terms are always diagonal, while the spin--orbit $\rm LS_{-}$-~\eqref{Eq:LS} and tensor $\rm T$-~\eqref{Eq:T} terms are generally non-diagonal. In the case of a symmetric composition the numerical multiplier before the spin--orbit $\rm LS_{-}$-term vanishes and it effectively drops out of the calculations while the tensor $\rm T$-term introduces only numerically insignificant mixing for orbitally excited (e.g. P-,D-wave) states with $J^{PC}=1^{--}, 2^{++}$ which can be neglected. In the case of an asymmetric composition the spin--orbit $\rm LS_{-}$-term plays a substantial role in mixing of the orbitally excited tetraquark states with the axialvector both diquark and antidiquark.
\par
The masses of the corresponding mixed states are the eigenvalues of the mixing matrices. We introduce the following notations:
\begin{widetext}
\begin{equation}
\label{Eq:not}
\begin{gathered}
M_{L=a, \, J=b} \equiv M_{a,b},
\\
M_{L=a, \, J=b}({S=c, \, S'=d}) \equiv M_{a,b}({c,d}),
\\		
\Delta M_{a,b}({c,d}) = \big[ M_{a,b}({c,d}) \big]_{\rm full} - \big[ M_{a,b}({c,d}) \big]_{\rm spin-ind.},
\end{gathered}
\end{equation}
\end{widetext}
\noindent
where $[M]_{\rm full}$ denotes the mass calculated with the complete quasipotential~\eqref{Eq:V}, and $[M]_{\rm spin-ind.}$ denotes the mass calculated with only its spin-independent part~\eqref{Eq:Va}. The mixing matrices then can be expressed as follows. 
\begin{itemize}

	\item For the P-wave states the masses for the given total momenta $J$ are eigenvalues of the matrices
	\begin{widetext}
	\begin{align}
	& J=1: \quad M_{1,1} = \, \lambda \!
	\begin{pmatrix}
		M_{1,1}(0,0) & \Delta M_{1,1}(0,1) & \Delta M_{1,1}(0,2) \\
		\Delta M_{1,1}(1,0) & M_{1,1}(1,1) & \Delta M_{1,1}(1,2) \\
		\Delta M_{1,1}(2,0) & \Delta M_{1,1}(2,1) & M_{1,1}(2,2)
	\end{pmatrix}, \nonumber
	\\
	& J=2: \quad M_{1,2} = \, \lambda \!
	\begin{pmatrix}
		M_{1,2}(1,1) & \Delta M_{1,2}(1,2) \\
		\Delta M_{1,2}(2,1) & M_{1,2}(2,2)
	\end{pmatrix}. \label{Eq:Pmix}
	\end{align}
	\end{widetext}
	
	\item Similarly, mixing matrices for the D-wave states
	\begin{widetext}
	\begin{align}
	& J=1: \quad M_{2,1} = \, \lambda \! 
	\begin{pmatrix}
		M_{2,1}(1,1) & \Delta M_{2,1}(1,2) \\
		\Delta M_{2,1}(2,1) & M_{2,1}(2,2)
	\end{pmatrix}, \nonumber
	\\
	& J=2: \quad M_{2,2} = \, \lambda \! 
	\begin{pmatrix}
		M_{2,2}(0,0) & \Delta M_{2,2}(0,1) & \Delta M_{2,2}(0,2) \\
		\Delta M_{2,2}(1,0) & M_{2,2}(1,1) & \Delta M_{2,2}(1,2) \\
		\Delta M_{2,2}(2,0) & \Delta M_{2,2}(2,1) & M_{2,2}(2,2)
	\end{pmatrix}, \label{Eq:Dmix}
	\\		
	& J=3: \quad M_{2,3} = \, \lambda \! 
	\begin{pmatrix}
		M_{2,3}(1,1) & \Delta M_{2,3}(1,2) \\
		\Delta M_{2,3}(2,1) & M_{2,3}(2,2)
	\end{pmatrix}. \nonumber
	\end{align}
	\end{widetext}
	
\end{itemize}
\par
The general algorithm of the calculations of the tetraquark masses is as follows. First, we calculate the masses and wave functions of the doubly heavy diquarks as the bound quark--quark states. It is done by solving Eq.~\eqref{Eq:Schr} with the quark--quark interaction quasipotential (see Eqs.~(5)-(15) in Ref.~\cite{Savch2022Sym}) numerically. The initial value of the relativistic reduced mass $\mu_{\rm R}$~\eqref{Eq:muR} is fixed to its nonrelativistic value and Eq.~\eqref{Eq:Schr} is solved as the Schr\"odinger equation (for details see Ref.~\cite{WolframSchr}). Then the masses of the bound states are found by the successive approximations method (for details see Ref.~\cite{Faustov1990eng}). The masses of the tetraquarks and their wave functions are calculated for the bound diquark--antidiquark states with the quasipotential~\eqref{Eq:V} using the same procedure with the obtained diquark masses as constituent masses.
\par
All free parameters of our model, such as the confinement potential mixing coefficient $\varepsilon$, anomalous chromomagnetic moment $\kappa$, parameter of the running coupling constant $\Lambda$, confining potential parameters $A, B$ and quark masses $m_{\rm c,b}$ were fixed previously and are taken from the study of the properties of mesons and baryons~\cite{Param1986, Param1990, Param1992, Param1995}. They are listed in Table~\ref{Tab:ParGen}. The diquark masses $M_{\rm cc, cb, bb}$ and the parameters of their form factors $\xi$ and $\zeta$ were already calculated earlier~\cite{Baryo2002, Tetra2007} and are given in Table~\ref{Tab:ParDi}.

\begin{table*}
\caption{Parameters of the model from Refs.~\cite{Param1986, Param1990, Param1992, Param1995}.\label{Tab:ParGen}}
\begin{ruledtabular}
\begin{tabular}{ccccccc}
$m_{\rm c}$, GeV & $m_{\rm b}$, GeV & $A$, GeV$^2$ & $B$, GeV & $\Lambda$, MeV & $\varepsilon$ & $\kappa$
\\
\hline
1.55 & 4.88 & 0.18 & $-$0.3 & 414 & $-$1 & $-$1
\end{tabular}
\end{ruledtabular}
\end{table*}

\begin{table*}
\caption{Masses $M_{\rm QQ'}$ and form factor parameters $\xi, \zeta$ of the diquarks from Refs.~\cite{Baryo2002, Tetra2007}. $\rm d$ is the axialvector (A) or scalar (S) diquark. $\rm [Q,Q']$ and $\rm \{Q,Q'\}$ denote combinations of quarks asymmetric and symmetric in flavor, respectively.\label{Tab:ParDi}}
\begin{ruledtabular}
\begin{tabular}{cccccccc}
\multirow{2}{*}{$\rm QQ'$} & \multirow{2}{*}{d} & \multicolumn{3}{c}{$\rm Q=c$} & \multicolumn{3}{c}{$\rm Q=b$}
\\
\cline{3-8}
 &  & $M_{\rm cQ'}$, MeV & $\xi$, GeV & $\zeta$, GeV$^2$ & $M_{\rm bQ'}$, MeV & $\xi$, GeV & $\zeta$, GeV$^2$
\\
\hline
$[\rm Q,c]$ & S &  &  &  & 6,519 & 1.50 & 0.59
\\ 
$\{ \rm Q,c \}$ & A & 3,226 & 1.30 & 0.42 & 6,526 & 1.50 & 0.59
\\
$\{ \rm Q,b \}$ & A & 6,526 & 1.50 & 0.59 & 9,778 & 1.30 & 1.60
\end{tabular}
\end{ruledtabular}
\end{table*}

\section{Masses of Asymmetric Fully Heavy Tetraquarks\label{Sec:Res}}

\par
The calculated mass spectra of ground states (1S) and orbitally and radially excited states (1P, 2S, 1D, 2P, 3S) of asymmetric fully heavy tetraquarks ($\rm cc\overline c\overline b$, $\rm bc\overline c\overline c$, $\rm cc\overline b\overline b$, $\rm bb\overline c\overline c$, $\rm bb\overline b\overline c$, $\rm cb\overline b\overline b$) are presented in Table~\ref{Tab:Res}. Note that the account of the finite size of the diquark weakens the one-gluon exchange potential thus increasing the predicted tetraquark mass. Masses of the ground states of all possible flavor compositions (i.e. three symmetric and six asymmetric), as well as the same excitations of symmetric fully heavy tetraquarks ($\rm cc\overline c\overline c$, $\rm cb\overline c\overline b$, $\rm bb\overline b\overline b$) have already been calculated in our previous papers~\cite{Savch2020PhysRevD, Savch2021Uni, Savch2022Sym}. Although, the mixing arising exclusively from the tensor term between the orbitally excited symmetric states was not taken into account since, as we just discussed in Sec.~\ref{Sec:Math}, in the symmetric case mixing can be discarded without loss of accuracy.
\par
As we discussed in Sec.~\ref{Sec:Model} a diquark can be either in scalar state with spin $S_{\rm d}=0$ or in axialvector state with $S_{\rm d}=1$. According to the Pauli principle the scalar diquark must contain quarks of different flavors while all diquarks can be in axialvector state. Therefore, the $\rm cc\overline b\overline b$, $\rm bb\overline c\overline c$ tetraquarks can contain only axialvector diquarks and antidiquarks while the $\rm cc\overline c\overline b$, $\rm bc\overline c\overline c$, $\rm bb\overline b\overline c$, $\rm cb\overline b\overline b$ tetraquarks can consist both from the axialvector diquark--antidiquark pair or a mixture of axialvector and scalar diquarks and antidiquarks. As a result, the $\rm cc\overline c\overline b$, $\rm bc\overline c\overline c$, $\rm bb\overline b\overline c$, $\rm cb\overline b\overline b$ tetraquarks have more possible states, an additional 12 mixed scalar--axialvector states are added to the 32 pure-axialvector states.
\par
It can be seen from Table~\ref{Tab:Res} that the predicted spectroscopy of the ground and excited states of tetraquarks within the diquark--antidiquark model is very rich. Nonetheless, not all of the predicted states can be reliably observed in experiments due to the fast fall-apart strong decays into two heavy mesons through the quark rearrangement. In order to determine states that are the most promising to live long enough to be observed, we can put a number of limitations. 
\begin{itemize}

	\item First, we consider diquarks only in the color-triplet state. This choice was explained in Sec.~\ref{Sec:Model} and is based on the following argument. In the color-antitriplet diquarks the internal quark--quark interaction potential is attractive thus the bound state is possible. Contrary the color-sextet diquarks have the repulsive internal interaction potential which makes the bound state unlikely to exist. Nevertheless, some researchers argue that the color-sextet diquarks can also contribute to the observable spectroscopy (for the discussion related to the asymmetric fully heavy tetraquarks in particular, see Refs.~\cite{QQQQ.ground:all.2018.2, QQQQ.ground;exc:cccc;bbbb;ccbb.2019.4, QQQQ.ground:all.exc:ccbb.2021.1, QQQQ.ground:all.2021.3, QQQQ.ground:all.2022.3, QQQQ.ground:all.2023.1}) and sometimes even dominate it (see Refs.~\cite{QQQQ.ground:all.exc:ccbb.2021.1, QQQQ.ground:all.2021.3, QQQQ.ground:all.2023.1} for the argument on the some ground states being dominated by the $(6 \times \overline 6)$-color configuration).
	
	\item Second, we consider only the ground state diquarks with excitations occurring only between the diquark and antidiquark. Such excitations lead to a larger separation between the diquark and antidiquark within the tetraquark, consequently increasing the mean distance between the heavy quarks of the diquark and antiquarks from the antidiquark. Such configuration reduces the probability of the rapid fall-apart strong decay process and, as a result, these states have more chances to be observed as resonances. Contrarily, excitations within the diquark and/or antidiquark increase the diquark mass and size thus providing a larger overlap between the diquark and antidiquark which enhances the rapid fall-apart strong decay processes.
	
\end{itemize}
\noindent
Additionally, suppression by the phase space, which is determined by the difference between the tetraquark mass and the meson pair decay threshold, increases chances of the state to be observed experimentally as a relatively narrow resonance (more on this topic in the following section). However, the excited states of the tetraquarks could be narrow despite the large phase space since it is necessary to overcome the suppression in the fall-apart process, either due to the centrifugal barrier for the orbital excitations or due to the presence of the nodes in the wave function of the radially excited state.

\begin{table*}
\caption{Masses $M_{\rm QQ'\overline Q\overline Q'}$ of the ground (1S) and excited (1P, 2S, 1D, 2P, 3S) states of the asymmetric fully heavy tetraquarks ($\rm cc\overline c\overline b$, $\rm bc\overline c\overline c$, $\rm cc\overline b\overline b$, $\rm bb\overline c\overline c$, $\rm bb\overline b\overline c$, $\rm cb\overline b\overline b$). $\rm d$ and $\rm \overline d'$ are the axialvector (A) or scalar (S) diquark and antidiquark, respectively. $n\rm L$ denotes the overall excitation. $n_{r}$ is the radial quantum number. $L$ is the relative orbital momentum between the diquark and antidiquark. $S$ is the total spin of the diquark--antidiquark system. $J^{P}$ is the total momentum-parity of the tetraquark. All masses are given in MeV.\label{Tab:Res}}
\begin{ruledtabular}
\begin{tabular}{ccccccccc}
$\rm d\overline d'$ & $n\rm L$ & $n_{r}$ & $L$ & $S$ & $J^{P}$ & $M_{\rm cc\overline c\overline b, bc\overline c\overline c}$ & $M_{\rm cc\overline b\overline b, bb\overline c\overline c}$ & $M_{\rm bb\overline b\overline c, cb\overline b\overline b}$
\\
\hline
\multirow{32}{*}{$\rm A\overline A$} & \multirow{3}{*}{1S} & \multirow{3}{*}{0} & \multirow{3}{*}{0} & 0 & $0^{+}$ & 9,606 & 12,848 & 16,102
\\
\cline{5-9}
 & & & & 1 & $1^{+}$ & 9,611 & 12,852 & 16,104
\\
\cline{5-9}
 & & & & 2 & $2^{+}$ & 9,620 & 12,859 & 16,108
\\
\cline{2-9}
 & \multirow{7}{*}{1P} & \multirow{7}{*}{0} & \multirow{7}{*}{1} & 1 & $0^{-}$ & 9,875 & 13,106 & 16,326
\\
\cline{5-9}
 & & & & 0 & \multirow{3}{*}{$1^{-}$} & 9,871 & 13,103 & 16,325
\\
 & & & & 1 & & 9,877 & 13,108 & 16,326
\\
 & & & & 2 & & 9,881 & 13,111 & 16,329
\\
\cline{5-9}
 & & & & 1 & \multirow{2}{*}{$2^{-}$} & 9,875 & 13,106 & 16,327
\\
 & & & & 2 & & 9,882 & 13,112 & 16,329
\\
\cline{5-9}
 & & & & 2 & $3^{-}$ & 9,881 & 13,110 & 16,330 
\\
\cline{2-9}
 & \multirow{3}{*}{2S} & \multirow{3}{*}{1} & \multirow{3}{*}{0} & 0 & $0^{+}$ & 10,063 & 13,282 & 16,481
\\
\cline{5-9}
 & & & & 1 & $1^{+}$ & 10,064 & 13,282 & 16,481
\\
\cline{5-9}
 & & & & 2 & $2^{+}$ & 10,064 & 13,283 & 16,481
\\
\cline{2-9}
 & \multirow{9}{*}{1D} & \multirow{9}{*}{0} & \multirow{9}{*}{2} & 2 & $0^{+}$ & 10,113 & 13,330 & 16,513
\\
\cline{5-9}
 & & & & 1 & \multirow{2}{*}{$1^{+}$} & 10,111 & 13,328 & 16,513
\\
 & & & & 2 & & 10,114 & 13,331 & 16,514 
\\
\cline{5-9}
 & & & & 0 & \multirow{3}{*}{$2^{+}$} & 10,108 & 13,324 & 16,513
\\
 & & & & 1 & & 10,113 & 13,330 & 16,514
\\
 & & & & 2 & & 10,117 & 13,334 & 16,515
\\
\cline{5-9}
 & & & & 1 & \multirow{2}{*}{$3^{+}$} & 10,111 & 13,327 & 16,515
\\
 & & & & 2 & & 10,116 & 13,332 & 16,516
\\
\cline{5-9}
 & & & & 2 & $4^{+}$ & 10,114 & 13,329 & 16,516 
\\
\cline{2-9}
 & \multirow{7}{*}{2P} & \multirow{7}{*}{1} & \multirow{7}{*}{1} & 1 & $0^{-}$ & 10,265 & 13,468 & 16,631
\\
\cline{5-9}
 & & & & 0 & \multirow{3}{*}{$1^{-}$} & 10,258 & 13,461 & 16,629
\\
 & & & & 1 & & 10,264 & 13,468 & 16,630
\\
 & & & & 2 & & 10,270 & 13,472 & 16,633
\\
\cline{5-9}
 & & & & 1 & \multirow{2}{*}{$2^{-}$} & 10,260 & 13,463 & 16,630
\\
 & & & & 2 & & 10,268 & 13,470 & 16,632
\\
\cline{5-9}
 & & & & 2 & $3^{-}$ & 10,263 & 13,466 & 16,631 
\\
\cline{2-9}
 & \multirow{3}{*}{3S} & \multirow{3}{*}{2} & \multirow{3}{*}{0} & 0 & $0^{+}$ & 10,442 & 13,629 & 16,765
\\
\cline{5-9}
 & & & & 1 & $1^{+}$ & 10,442 & 13,629 & 16,765
\\
\cline{5-9}
 & & & & 2 & $2^{+}$ & 10,440 & 13,628 & 16,764
\\
\hline
\multirow{12}{*}{$\rm S\overline A$, $\rm A\overline S$} & 1S & 0 & 0 & \multirow{12}{*}{1} & $1^{+}$ & 9,608 &  & 16,099 
\\
\cline{2-4}
\cline{6-7}
\cline{9-9}
 & \multirow{3}{*}{1P} & \multirow{3}{*}{0} & \multirow{3}{*}{1} &  & $0^{-}$ & 9,873 &  & 16,320 
\\
\cline{6-7}
\cline{9-9}
 &  &  &  &  & $1^{-}$ & 9,872 &  & 16,321 
\\
\cline{6-7}
\cline{9-9}
 &  &  &  &  & $2^{-}$ & 9,871 &  & 16,322 
\\
\cline{2-4}
\cline{6-7}
\cline{9-9}
 & 2S & 1 & 0 &  & $1^{+}$ & 10,057 &  & 16,474 
\\
\cline{2-4}
\cline{6-7}
\cline{9-9}
 & \multirow{3}{*}{1D} & \multirow{3}{*}{0} & \multirow{3}{*}{2} &  & $1^{+}$ & 10,108 &  & 16,507 
\\
\cline{6-7}
\cline{9-9}
 &  &  &  &  & $2^{+}$ & 10,107 &  & 16,508 
\\
\cline{6-7}
\cline{9-9}
 &  &  &  &  & $3^{+}$ & 10,105 &  & 16,509 
\\
\cline{2-4}
\cline{6-7}
\cline{9-9}
 & \multirow{3}{*}{2P} & \multirow{3}{*}{1} & \multirow{3}{*}{1} &  & $0^{-}$ & 10,262 &  & 16,624 
\\
\cline{6-7}
\cline{9-9}
 &  &  &  &  & $1^{-}$ & 10,260 &  & 16,624 
\\
\cline{6-7}
\cline{9-9}
 &  &  &  &  & $2^{-}$ & 10,254 &  & 16,624 
\\
\cline{2-4}
\cline{6-7}
\cline{9-9}
 & 3S & 2 & 0 & & $1^{+}$ & 10,434 &  & 16,758 
\end{tabular}
\end{ruledtabular}
\end{table*}

\section{Threshold Analysis\label{Sec:Thr}}

\par
In the previous section we have calculated the masses of asymmetric fully heavy tetraquarks. Now we need to understand which of the predicted tetraquark states could be observed experimentally, i.e. could appear as narrow resonances in different decay modes. Note that whether the state is truly a resonance is a difficult question requiring additional complex analysis. Such analysis is beyond the scope of the present paper. In the following we only identify states which could be observed as resonances. The probability rate of different types of decays is as follows.
\begin{itemize}

	\item The most probable is the fall-apart strong decay, where the quarks and antiquarks from the initial tetraquark rearrange into two mesons
	\begin{equation}
	\rm Q_{1}Q_{2}\overline Q_{3}\overline Q_{4} \longrightarrow Q_{1}\overline Q_{3} \ + \ Q_{2}\overline Q_{4}. \label{decay:fallapart}
	\end{equation}
	\noindent
	If energetically possible and isn't forbidden by the quantum numbers, this is the main decay channel. As the state can decay very rapidly, it could be observed as a broad resonance in this channel.
	
	\item Then there is the strong decay due to the heavy quark--antiquark annihilation into gluons ($\rm g$), followed by their hadronization into lighter hadrons ($\rm h_{i}$)
	\begin{equation}
	\rm Q_{1}Q_{2}\overline Q_{3}\overline Q_{4} \longrightarrow g \ + \ g \ +  ... \longrightarrow h_{1} \ + \ h_{2} \ + ... . \label{decay:annihil}
	\end{equation}
	\noindent
	Such processes are strongly suppressed by the Okubo-Zweig-Iizuki rule. Nevertheless, such decays are always possible for the fully heavy tetraquarks which contain at least one quark--antiquark pair of the same flavor (we have only two heavy quark flavors). Thus, if the fall-apart strong decay is forbidden the state could be observed as narrow in this decay channel.
	
	\item Even less probable is the radiative decay which is determined by the radiative transition from the higher to the lower excitation of the tetraquark state
	\begin{equation}
	\rm Q_{1}Q_{2}\overline Q_{3}\overline Q_{4} \longrightarrow Q_{1}Q_{2}\overline Q_{3}\overline Q_{4} \ + \ \gamma. \label{decay:radiat}
	\end{equation}
	\noindent
	These processes are suppressed by the significantly lower coupling constant of the electromagnetic interaction $\alpha$ ($\alpha \ll \alpha_{\rm S}$). Thus, if the fall-apart strong decay is forbidden and and gluon-annihilation is significantly suppressed than the tetraquark could be observed as a very narrow state in this decay channel.
	
	\item In fact, the least probable is the weak decay which can be completely neglected in this discussion due to the smallness of the weak coupling constant $\alpha_{\rm W}$ ($\alpha_{\rm W} \ll \alpha \ll \alpha_{\rm S}$). However, the possible weak decay channels of the $\rm cc\overline b\overline b$, $\rm bb\overline c\overline c$ tetraquark in the ground state are discussed in Ref.~\cite{QQQQ:ccbb.2019.2}.
	
\end{itemize}
\par
Therefore, in order to analyze the possibility of the experimental observations it is useful to compare the calculated tetraquark masses with the fall-apart strong decay thresholds, i.e. masses of pairs of mesons composed of the same quarks and antiquarks as the initial tetraquark. This difference determines the phase space and, thus, the probability of the corresponding fall-apart strong decays. Hence, two cases can be distinguished. 
\begin{itemize}

	\item If a mass of the tetraquark exceeds the mass of the meson pair (energetically possible) and total momentum--parity $J^{P}$ of the initial and final systems coincides (allowed by the quantum numbers), then the fall-apart strong decay~\eqref{decay:fallapart} is possible and it is the main channel. As we have just discussed, generally such states could be observed as broad resonances. Note that the closer the tetraquark mass is to the threshold the narrower the state can be. 
	
	\item If a mass of the tetraquark lies below the corresponding threshold then the fall-apart strong decay~\eqref{decay:fallapart} is forbidden and the main channels will be either gluon-annihilation~\eqref{decay:annihil} or radiative decays~\eqref{decay:radiat}, if allowed. Such processes are both strongly suppressed and, as a result, these tetraquark states are narrow and could be observed experimentally in other decay channels into hadrons composed of lighter quarks and antiquarks, or into the lower tetraquark excitation and a photon.
	
\end{itemize}
\par
Note than an excited tetraquark can fall-apart into a meson pair, allowed by the total momentum--parity conservation, either in the S-wave with the orbital momentum between these mesons $L_{\rm thr.} = 0$, or in the higher wave with $L_{\rm thr.} > 0$. In the former case the final state contains orbitally excited mesons with heavier masses. In the latter case the final mesons have lower masses due to the lower orbital excitations. However, such decays are highly suppressed by the additional factor $(\mathbf{p}^{2}/M^{2})^{L_{\rm thr.}}$, where $\mathbf{p}$ is the relative momentum. The estimates of these suppression factors for the fully heavy tetraquarks give values less than $10^{-2}$. Thus in this paper we consider only the S-wave fall-apart strong decays.
\par
In Tables~\ref{Tab:Thrcccb}-\ref{Tab:Thrbbbc} we compare our calculations for the asymmetric fully heavy tetraquarks masses from Table~\ref{Tab:Res} with the meson--meson fall-apart strong decay thresholds. As we just discussed, the values of the mass separation from the threshold $\Delta$ which determines the phase space volume are of special interest
\begin{equation}
\label{Eq:Delta}
\Delta = M_{\rm QQ'\overline Q\overline Q'} - M_{\rm thr.},
\end{equation}
where $M_{\rm QQ'\overline Q\overline Q'}$ is the tetraquark mass and $M_{\rm thr.}$ is the threshold for the decay into a meson pair which is the sum of the meson masses. We are interested in the most probable decay modes for each state. The rate of the fall-apart strong decay is governed by the phase space volume. The main principle is the following: the smaller the phase space is, the less energetically favorable the decay is. Consequently, at any given time, the probability of transition to the final state (i.e. decay) is small. Thus the main decay channel being the most probable decay mode corresponds to the largest of possible values of $\Delta$: $\Delta_{\rm max}$. Therefore, in Tables~\ref{Tab:Thrcccb}-\ref{Tab:Thrbbbc} we compare the masses of tetraquarks only with the lowest fall-apart strong decay thresholds, as
\begin{equation*}
\label{Eq:ThrProb}
[M_{\rm thr.}]_{\rm min} \leftrightarrow \Delta_{\rm max} \leftrightarrow \text{more probable decay mode}.
\end{equation*}

\begin{table*}
\caption{Masses $M_{\rm QQ'\overline Q\overline Q'}$ of the ground (1S) and excited (1P, 2S, 1D, 2P, 3S) states of the triple charmed and bottom ($\rm cc\overline c\overline b$, $\rm bc\overline c\overline c$) fully heavy tetraquark, composed from the axialvector and/or scalar diquarks and the corresponding meson--meson fall-apart strong decay thresholds. $\rm d$ and $\rm \overline d'$ are the axialvector (A) or scalar (S) diquark and antidiquark, respectively. $n\rm L$ denotes the overall excitation. $S$ is the total spin of the diquark--antidiquark system. $J^{P}$ is the total momentum-parity of the tetraquark. $M_{\rm thr.}$ is the corresponding meson–-meson threshold~\cite{PDG2022, Regge2011meson}. $\Delta_{\rm max}$ is the difference between the tetraquark mass and lowest fall-apart threshold $\Delta_{\rm max} = M_{\rm QQ'\overline Q\overline Q'} - [M_{\rm thr.}]_{\rm min}$. All masses are given in MeV. States with the $\Delta_{\rm max} \leq 100$ MeV are additionally highlighted in violet and bold. States with the $\Delta_{\rm max} < 0$ are additionally highlighted in red and bold. The lowest thresholds, containing the $B_{c}^{\pm}(1 \prescript{1}{}{S}_{0})$-meson, are additionally highlighted in bold.\label{Tab:Thrcccb}}
\begin{ruledtabular}
\begin{tabular}{ccccccccc}
$\rm QQ'\overline Q\overline Q'$ & $\rm d\overline d'$ & $n\rm L$ & $S$ & $J^{P}$ & $M_{\rm QQ'\overline Q\overline Q'}$ & $M_{\rm thr.}$ & $\Delta_{\rm max}$ & Meson pair
\\
\hline
\multirow{44}{*}{\makecell{$\rm cc\overline c\overline b$, \\ $\rm bc\overline c\overline c$}} & \multirow{32}{*}{$\rm A\overline A$} & \multirow{3}{*}{1S} & 0 & $0^{+}$ & 9,606 & 9,258 & 348 & $\rm \boldsymbol \eta_{c}\mathbf{(1S) \ B_{c}^{\pm}(1 \prescript{1}{}{S}_{0})}$
\\
\cline{4-9}
 & & & 1 & $1^{+}$ & 9,611 & 9,317 & 294 & $\rm \eta_{c}(1S) \ B_{c}^{\pm}(1 \prescript{3}{}{S}_{1})$
\\
\cline{4-9}
 & & & 2 & $2^{+}$ & 9,620 & 9,430 & 190 & $\rm J/\psi(1S) \ B_{c}^{\pm}(1 \prescript{3}{}{S}_{1})$
\\
\cline{3-9}
 & & \multirow{7}{*}{1P} & 1 & $0^{-}$ & 9,875 & 9,683 & 192 & $\rm \eta_{c}(1S) \ B_{c}^{\pm}(1 \prescript{3}{}{P}_{0})$
\\
\cline{4-9}
 & & & 0 & \multirow{3}{*}{$1^{-}$} & 9,871 & \multirow{3}{*}{9,727} & 144 & \multirow{3}{*}{$\rm \eta_{c}(1S) \ B_{c}^{\pm}(1 \prescript{ }{}{P}_{1})$}
\\
 & & & 1 & & 9,877 & & 150 & 
\\
 & & & 2 & & 9,881 & & 154 & 
\\
\cline{4-9}
 & & & 1 & \multirow{2}{*}{$2^{-}$} & 9,875 & \multirow{2}{*}{9,745} & 130 & \multirow{2}{*}{$\rm \eta_{c}(1S) \ B_{c}^{\pm}(1 \prescript{3}{}{P}_{2})$}
\\
 & & & 2 & & 9,882 & & 137 & 
\\
\cline{4-9}
 & & & 2 & \textcolor{violet}{$\mathbf{3^{-}}$} & \textcolor{violet}{\textbf{9,881}} & 9,858 & \textcolor{violet}{\textbf{23}} & $\rm J/\psi(1S) \ B_{c}^{\pm}(1 \prescript{3}{}{P}_{2})$
\\
\cline{3-9}
 & & \multirow{3}{*}{2S} & 0 & $0^{+}$ & 10,063 & 9,258 & 805 & $\rm \boldsymbol \eta_{c}\mathbf{(1S) \ B_{c}^{\pm}(1 \prescript{1}{}{S}_{0})}$
\\
\cline{4-9}
 & & & 1 & $1^{+}$ & 10,064 & 9,317 & 747 & $\rm \eta_{c}(1S) \ B_{c}^{\pm}(1 \prescript{3}{}{S}_{1})$
\\
\cline{4-9}
 & & & 2 & $2^{+}$ & 10,064 & 9,430 & 634 & $\rm J/\psi(1S) \ B_{c}^{\pm}(1 \prescript{3}{}{S}_{1})$
\\
\cline{3-9} 
 & & \multirow{9}{*}{1D} & 2 & $0^{+}$ & 10,113 & 9,258 & 855 & $\rm \boldsymbol \eta_{c}\mathbf{(1S) \ B_{c}^{\pm}(1 \prescript{1}{}{S}_{0})}$
\\
\cline{4-9}
 & & & 1 & \multirow{2}{*}{$1^{+}$} & 10,111 & \multirow{2}{*}{9,317} & 794 & \multirow{2}{*}{$\rm \eta_{c}(1S) \ B_{c}^{\pm}(1 \prescript{3}{}{S}_{1})$}
\\
 & & & 2 & & 10,114 & & 797 & 
\\
\cline{4-9}
 & & & 0 & \multirow{3}{*}{$2^{+}$} & 10,108 & \multirow{3}{*}{9,430} & 678 & \multirow{3}{*}{$\rm J/\psi(1S) \ B_{c}^{\pm}(1 \prescript{3}{}{S}_{1})$}
\\
 & & & 1 & & 10,113 & & 683 & 
\\
 & & & 2 & & 10,117 & & 687 & 
\\
\cline{4-9}
 & & & 1 & \multirow{2}{*}{\textcolor{violet}{$\mathbf{3^{+}}$}} & \textcolor{violet}{\textbf{10,111}} & \multirow{2}{*}{10,013} & \textcolor{violet}{\textbf{98}} & \multirow{2}{*}{$\rm \eta_{c}(1S) \ B_{c}^{\pm}(1 \prescript{3}{}{D}_{3})$}
\\
 & & & 2 & & 10,116 & & 103 & 
\\
\cline{4-9}
 & & & 2 & \textcolor{red}{$\mathbf{4^{+}}$} & \textcolor{red}{\textbf{10,114}} & 10,126 & \textcolor{red}{\textbf{-12}} & $\rm J/\psi(1S) \ B_{c}^{\pm}(1 \prescript{3}{}{D}_{3})$
\\
\cline{3-9}
 & & \multirow{7}{*}{2P} & 1 & $0^{-}$ & 10,265 & 9,683 & 582 & $\rm \eta_{c}(1S) \ B_{c}^{\pm}(1 \prescript{3}{}{P}_{0})$
\\
\cline{4-9}
 & & & 0 & \multirow{3}{*}{$1^{-}$} & 10,258 & \multirow{3}{*}{9,727} & 531 & \multirow{3}{*}{$\rm \eta_{c}(1S) \ B_{c}^{\pm}(1 \prescript{ }{}{P}_{1})$}
\\
 & & & 1 & & 10,264 & & 537 & 
\\
 & & & 2 & & 10,270 & & 543 & 
\\
\cline{4-9}
 & & & 1 & \multirow{2}{*}{$2^{-}$} & 10,260 & \multirow{2}{*}{9,745} & 515 & \multirow{2}{*}{$\rm \eta_{c}(1S) \ B_{c}^{\pm}(1 \prescript{3}{}{P}_{2})$}
\\
 & & & 2 & & 10,268 & & 523 & 
\\
\cline{4-9}
 & & & 2 & $3^{-}$ & 10,263 & 9,858 & 405 & $\rm J/\psi(1S) \ B_{c}^{\pm}(1 \prescript{3}{}{P}_{2})$
\\
\cline{3-9}
 & & \multirow{3}{*}{3S} & 0 & $0^{+}$ & 10,442 & 9,258 & 1,184 & $\rm \boldsymbol \eta_{c}\mathbf{(1S) \ B_{c}^{\pm}(1 \prescript{1}{}{S}_{0})}$
\\
\cline{4-9}
 & & & 1 & $1^{+}$ & 10,442 & 9,317 & 1,125 & $\rm \eta_{c}(1S) \ B_{c}^{\pm}(1 \prescript{3}{}{S}_{1})$
\\
\cline{4-9}
 & & & 2 & $2^{+}$ & 10,440 & 9,430 & 1,010 & $\rm J/\psi(1S) \ B_{c}^{\pm}(1 \prescript{3}{}{S}_{1})$
\\
\cline{2-9}
 & \multirow{12}{*}{$\rm S\overline A$, $\rm A\overline S$} & 1S & \multirow{12}{*}{1} & $1^{+}$ & 9,608 & 9,317 & 291 & $\rm \eta_{c}(1S) \ B_{c}^{\pm}(1 \prescript{3}{}{S}_{1})$
\\
\cline{3-3}
\cline{5-9}
 & & \multirow{3}{*}{1P} & & $0^{-}$ & 9,873 & 9,683 & 190 & $\rm \eta_{c}(1S) \ B_{c}^{\pm}(1 \prescript{3}{}{P}_{0})$
\\
\cline{5-9}
 & & & & $1^{-}$ & 9,872 & 9,727 & 145 & $\rm \eta_{c}(1S) \ B_{c}^{\pm}(1 \prescript{ }{}{P}_{1})$
\\
\cline{5-9}
 & & & & $2^{-}$ & 9,871 & 9,745 & 126 & $\rm \eta_{c}(1S) \ B_{c}^{\pm}(1 \prescript{3}{}{P}_{2})$
\\
\cline{3-3}
\cline{5-9}
 & & 2S & & $1^{+}$ & 10,057 & 9,317 & 740 & $\rm \eta_{c}(1S) \ B_{c}^{\pm}(1 \prescript{3}{}{S}_{1})$
\\
\cline{3-3}
\cline{5-9}
 & & \multirow{3}{*}{1D} & & $1^{+}$ & 10,108 & 9,317 & 791 & $\rm \eta_{c}(1S) \ B_{c}^{\pm}(1 \prescript{3}{}{S}_{1})$
\\
\cline{5-9}
 & & & & $2^{+}$ & 10,107 & 9,430 & 677 & $\rm J/\psi(1S) \ B_{c}^{\pm}(1 \prescript{3}{}{S}_{1})$
\\
\cline{5-9}
 & & & & \textcolor{violet}{$\mathbf{3^{+}}$} & \textcolor{violet}{\textbf{10,105}} & 10,013 & \textcolor{violet}{\textbf{92}} & $\rm \eta_{c}(1S) \ B_{c}^{\pm}(1 \prescript{3}{}{D}_{3})$
\\
\cline{3-3}
\cline{5-9}
 & & \multirow{3}{*}{2P} & & $0^{-}$ & 10,262 & 9,683 & 579 & $\rm \eta_{c}(1S) \ B_{c}^{\pm}(1 \prescript{3}{}{P}_{0})$
\\
\cline{5-9}
 & & & & $1^{-}$ & 10,260 & 9,727 & 533 & $\rm \eta_{c}(1S) \ B_{c}^{\pm}(1 \prescript{ }{}{P}_{1})$
\\
\cline{5-9}
 & & & & $2^{-}$ & 10,254 & 9,745 & 509 & $\rm \eta_{c}(1S) \ B_{c}^{\pm}(1 \prescript{3}{}{P}_{2})$
\\
\cline{3-3}
\cline{5-9}
 & & 3S & & $1^{+}$ & 10,434 & 9,317 & 1,117 & $\rm \eta_{c}(1S) \ B_{c}^{\pm}(1 \prescript{3}{}{S}_{1})$
\end{tabular}
\end{ruledtabular}
\end{table*}

\begin{table*}
\caption{Same as in Table~\ref{Tab:Thrcccb}, but for the double charmed and double bottom ($\rm cc\overline b\overline b$, $\rm bb\overline c\overline c$) fully heavy tetraquark, composed from the axialvector diquarks. \label{Tab:Thrccbb}}
\begin{ruledtabular}
\begin{tabular}{ccccccccc}
$\rm QQ'\overline Q\overline Q'$ & $\rm d\overline d'$ & $n\rm L$ & $S$ & $J^{P}$ & $M_{\rm QQ'\overline Q\overline Q'}$ & $M_{\rm thr.}$ & $\Delta_{\rm max}$ & Meson pair
\\
\hline
\multirow{32}{*}{\makecell{$\rm cc\overline b\overline b$, \\ $\rm bb\overline c\overline c$}} & \multirow{32}{*}{$\rm A\overline A$} & \multirow{3}{*}{1S} & 0 & $0^{+}$ & 12,848 & 12,549 & 299 & $\rm \mathbf{B_{c}^{\pm}(1 \prescript{1}{}{S}_{0}) \ B_{c}^{\pm}(1 \prescript{1}{}{S}_{0})}$
\\
\cline{4-9}
 & & & 1 & $1^{+}$ & 12,852 & 12,607 & 245 & $\rm B_{c}^{\pm}(1 \prescript{1}{}{S}_{0}) \ B_{c}^{\pm}(1 \prescript{3}{}{S}_{1})$
\\
\cline{4-9}
 & & & 2 & $2^{+}$ & 12,859 & 12,666 & 193 & $\rm B_{c}^{\pm}(1 \prescript{3}{}{S}_{1}) \ B_{c}^{\pm}(1 \prescript{3}{}{S}_{1})$
\\
\cline{3-9}
 & & \multirow{7}{*}{1P} & 1 & $0^{-}$ & 13,106 & 12,973 & 133 & $\rm B_{c}^{\pm}(1 \prescript{1}{}{S}_{0}) \ B_{c}^{\pm}(1 \prescript{3}{}{P}_{0})$
\\
\cline{4-9}
 & & & 0 & \textcolor{violet}{\multirow{3}{*}{$\mathbf{1^{-}}$}} & \textcolor{violet}{\textbf{13,103}} & \multirow{3}{*}{13,017} & \textcolor{violet}{\textbf{86}} & \multirow{3}{*}{$\rm B_{c}^{\pm}(1 \prescript{1}{}{S}_{0}) \ B_{c}^{\pm}(1 \prescript{ }{}{P}_{1})$}
\\
 & & & 1 & & \textcolor{violet}{\textbf{13,108}} & & \textcolor{violet}{\textbf{91}} &
\\
 & & & 2 & & \textcolor{violet}{\textbf{13,111}} & & \textcolor{violet}{\textbf{94}} &
\\
\cline{4-9}
 & & & 1 & \textcolor{violet}{\multirow{2}{*}{$\mathbf{2^{-}}$}} & \textcolor{violet}{\textbf{13,106}} & \multirow{2}{*}{13,035} & \textcolor{violet}{\textbf{71}} & \multirow{2}{*}{$\rm B_{c}^{\pm}(1 \prescript{1}{}{S}_{0}) \ B_{c}^{\pm}(1 \prescript{3}{}{P}_{2})$}
\\
 & & & 2 & & \textcolor{violet}{\textbf{13,112}} & & \textcolor{violet}{\textbf{77}} &
\\
\cline{4-9}
 & & & 2 & \textcolor{violet}{$\mathbf{3^{-}}$} & \textcolor{violet}{\textbf{13,110}} & 13,094 & \textcolor{violet}{\textbf{16}} & $\rm B_{c}^{\pm}(1 \prescript{3}{}{S}_{1}) \ B_{c}^{\pm}(1 \prescript{3}{}{P}_{2})$
\\
\cline{3-9}
 & & \multirow{3}{*}{2S} & 0 & $0^{+}$ & 13,282 & 12,549 & 733 & $\rm \mathbf{B_{c}^{\pm}(1 \prescript{1}{}{S}_{0}) \ B_{c}^{\pm}(1 \prescript{1}{}{S}_{0})}$
\\
\cline{4-9}
 & & & 1 & $1^{+}$ & 13,282 & 12,607 & 675 & $\rm B_{c}^{\pm}(1 \prescript{1}{}{S}_{0}) \ B_{c}^{\pm}(1 \prescript{3}{}{S}_{1})$
\\
\cline{4-9}
 & & & 2 & $2^{+}$ & 13,283 & 12,666 & 617 & $\rm B_{c}^{\pm}(1 \prescript{3}{}{S}_{1}) \ B_{c}^{\pm}(1 \prescript{3}{}{S}_{1})$
\\
\cline{3-9}
 & & \multirow{9}{*}{1D} & 2 & $0^{+}$ & 13,330 & 12,549 & 781 & $\rm \mathbf{B_{c}^{\pm}(1 \prescript{1}{}{S}_{0}) \ B_{c}^{\pm}(1 \prescript{1}{}{S}_{0})}$
\\
\cline{4-9}
 & & & 1 & \multirow{2}{*}{$1^{+}$} & 13,328 & \multirow{2}{*}{12,607} & 721 & \multirow{2}{*}{$\rm B_{c}^{\pm}(1 \prescript{1}{}{S}_{0}) \ B_{c}^{\pm}(1 \prescript{3}{}{S}_{1})$}
\\
 & & & 2 & & 13,331 & & 724 & 
\\
\cline{4-9}
 & & & 0 & \multirow{3}{*}{$2^{+}$} & 13,324 & \multirow{3}{*}{12,666} & 658 & \multirow{3}{*}{$\rm B_{c}^{\pm}(1 \prescript{3}{}{S}_{1}) \ B_{c}^{\pm}(1 \prescript{3}{}{S}_{1})$}
\\
 & & & 1 & & 13,330 & & 664 & 
\\
 & & & 2 & & 13,334 & & 668 & 
\\
\cline{4-9}
 & & & 1 & \textcolor{violet}{\multirow{2}{*}{$\mathbf{3^{+}}$}} & \textcolor{violet}{\textbf{13,327}} & \multirow{2}{*}{13,303} & \textcolor{violet}{\textbf{24}} & \multirow{2}{*}{$\rm B_{c}^{\pm}(1 \prescript{1}{}{S}_{0}) \ B_{c}^{\pm}(1 \prescript{3}{}{D}_{3})$}
\\
 & & & 2 & & \textcolor{violet}{\textbf{13,332}} & & \textcolor{violet}{\textbf{29}} & 
\\
\cline{4-9}
 & & & 2 & \textcolor{red}{$\mathbf{4^{+}}$} & \textcolor{red}{\textbf{13,329}} & 13,362 & \textcolor{red}{\textbf{-33}} & $\rm B_{c}^{\pm}(1 \prescript{3}{}{S}_{1}) \ B_{c}^{\pm}(1 \prescript{3}{}{D}_{3})$
\\
\cline{3-9}
 & & \multirow{7}{*}{2P} & 1 & $0^{-}$ & 13,468 & 12,973 & 495 & $\rm B_{c}^{\pm}(1 \prescript{1}{}{S}_{0}) \ B_{c}^{\pm}(1 \prescript{3}{}{P}_{0})$
\\
\cline{4-9}
 & & & 0 & \multirow{3}{*}{$1^{-}$} & 13,461 & \multirow{3}{*}{13,017} & 444 & \multirow{3}{*}{$\rm B_{c}^{\pm}(1 \prescript{1}{}{S}_{0}) \ B_{c}^{\pm}(1 \prescript{ }{}{P}_{1})$}
\\
 & & & 1 & & 13,468 & & 451 &
\\
 & & & 2 & & 13,472 & & 455 &
\\
\cline{4-9}
 & & & 1 & \multirow{2}{*}{$2^{-}$} & 13,463 & \multirow{2}{*}{13,035} & 428 & \multirow{2}{*}{$\rm B_{c}^{\pm}(1 \prescript{1}{}{S}_{0}) \ B_{c}^{\pm}(1 \prescript{3}{}{P}_{2})$}
\\
 & & & 2 & & 13,470 & & 435 &
\\
\cline{4-9}
 & & & 2 & $3^{-}$ & 13,466 & 13,094 & 372 & $\rm B_{c}^{\pm}(1 \prescript{3}{}{S}_{1}) \ B_{c}^{\pm}(1 \prescript{3}{}{P}_{2})$
\\
\cline{3-9}
 & & \multirow{3}{*}{3S} & 0 & $0^{+}$ & 13,629 & 12,549 & 1,080 & $\rm \mathbf{B_{c}^{\pm}(1 \prescript{1}{}{S}_{0}) \ B_{c}^{\pm}(1 \prescript{1}{}{S}_{0})}$
\\
\cline{4-9}
 & & & 1 & $1^{+}$ & 13,629 & 12,607 & 1,022 & $\rm B_{c}^{\pm}(1 \prescript{1}{}{S}_{0}) \ B_{c}^{\pm}(1 \prescript{3}{}{S}_{1})$
\\
\cline{4-9}
 & & & 2 & $2^{+}$ & 13,628 & 12,666 & 962 & $\rm B_{c}^{\pm}(1 \prescript{3}{}{S}_{1}) \ B_{c}^{\pm}(1 \prescript{3}{}{S}_{1})$
\end{tabular}
\end{ruledtabular}
\end{table*}

\begin{table*}
\caption{Same as in Table~\ref{Tab:Thrcccb}, but for the triple bottom and charmed ($\rm bb\overline b\overline c$, $\rm cb\overline b\overline b$) fully heavy tetraquark, composed from the axialvector and/or scalar diquarks.\label{Tab:Thrbbbc}}
\begin{ruledtabular}
\begin{tabular}{ccccccccc}
$\rm QQ'\overline Q\overline Q'$ & $\rm d\overline d'$ & $n\rm L$ & $S$ & $J^{P}$ & $M_{\rm QQ'\overline Q\overline Q'}$ & $M_{\rm thr.}$ & $\Delta_{\rm max}$ & Meson pair
\\
\hline
\multirow{44}{*}{\makecell{$\rm bb\overline b\overline c$, \\ $\rm cb\overline b\overline b$}} & \multirow{32}{*}{$\rm A\overline A$} & \multirow{3}{*}{1S} & 0 & $0^{+}$ & 16,102 & 15,673 & 429 & $\rm \boldsymbol \eta_{b}\mathbf{(1S) \ B_{c}^{\pm}(1 \prescript{1}{}{S}_{0})}$
\\
\cline{4-9}
 & & & 1 & $1^{+}$ & 16,104 & 15,732 & 372 & $\rm \eta_{b}(1S) \ B_{c}^{\pm}(1 \prescript{3}{}{S}_{1})$
\\
\cline{4-9}
 & & & 2 & $2^{+}$ & 16,108 & 15,793 & 315 & $\Upsilon(1S) \rm \ B_{c}^{\pm}(1 \prescript{3}{}{S}_{1})$
\\
\cline{3-9}
 & & \multirow{7}{*}{1P} & 1 & $0^{-}$ & 16,326 & 16,098 & 228 & $\rm \eta_{b}(1S) \ B_{c}^{\pm}(1 \prescript{3}{}{P}_{0})$
\\
\cline{4-9}
 & & & 0 & \multirow{3}{*}{$1^{-}$} & 16,325 & \multirow{3}{*}{16,142} & 183 & \multirow{3}{*}{$\rm \eta_{b}(1S) \ B_{c}^{\pm}(1 \prescript{ }{}{P}_{1})$}
\\
 & & & 1 & & 16,326 & & 184 & 
\\
 & & & 2 & & 16,329 & & 187 & 
\\
\cline{4-9}
 & & & 1 & \multirow{2}{*}{$2^{-}$} & 16,327 & \multirow{2}{*}{16,160} & 167 & \multirow{2}{*}{$\rm \eta_{b}(1S) \ B_{c}^{\pm}(1 \prescript{3}{}{P}_{2})$}
\\
 & & & 2 & & 16,329 & & 169 & 
\\
\cline{4-9}
 & & & 2 & $3^{-}$ & 16,330 & 16,221 & 109 & $\Upsilon(1S) \rm \ B_{c}^{\pm}(1 \prescript{3}{}{P}_{2})$
\\
\cline{3-9}
 & & \multirow{3}{*}{2S} & 0 & $0^{+}$ & 16,481 & 15,673 & 808 & $\rm \boldsymbol \eta_{b}\mathbf{(1S) \ B_{c}^{\pm}(1 \prescript{1}{}{S}_{0})}$
\\
\cline{4-9}
 & & & 1 & $1^{+}$ & 16,481 & 15,732 & 749 & $\rm \eta_{b}(1S) \ B_{c}^{\pm}(1 \prescript{3}{}{S}_{1})$
\\
\cline{4-9}
 & & & 2 & $2^{+}$ & 16,481 & 15,793 & 688 & $\Upsilon(1S) \rm \ B_{c}^{\pm}(1 \prescript{3}{}{S}_{1})$
\\
\cline{3-9}
 & & \multirow{9}{*}{1D} & 2 & $0^{+}$ & 16,513 & 15,673 & 840 & $\rm \boldsymbol \eta_{b}\mathbf{(1S) \ B_{c}^{\pm}(1 \prescript{1}{}{S}_{0})}$
\\
\cline{4-9}
 & & & 1 & \multirow{2}{*}{$1^{+}$} & 16,513 & \multirow{2}{*}{15,732} & 781 & \multirow{2}{*}{$\rm \eta_{b}(1S) \ B_{c}^{\pm}(1 \prescript{3}{}{S}_{1})$}
\\
 & & & 2 & & 16,514 & & 782 & 
\\
\cline{4-9}
 & & & 0 & \multirow{3}{*}{$2^{+}$} & 16,513 &  \multirow{3}{*}{15,793} & 720 &  \multirow{3}{*}{$\Upsilon(1S) \rm \ B_{c}^{\pm}(1 \prescript{3}{}{S}_{1})$}
\\
 & & & 1 & & 16,514 & & 721 & 
\\
 & & & 2 & & 16,515 & & 722 & 
\\
\cline{4-9}
 & & & 1 & \multirow{2}{*}{\textcolor{violet}{$\mathbf{3^{+}}$}} & \textcolor{violet}{\textbf{16,515}} & \multirow{2}{*}{16,428} & \textcolor{violet}{\textbf{87}} & \multirow{2}{*}{$\rm \eta_{b}(1S) \ B_{c}^{\pm}(1 \prescript{3}{}{D}_{3})$}
\\
 & & & 2 & & \textcolor{violet}{\textbf{16,516}} & & \textcolor{violet}{\textbf{88}} & 
\\
\cline{4-9}
 & & & 2 & \textcolor{violet}{$\mathbf{4^{+}}$} & \textcolor{violet}{\textbf{16,516}} & 16,489 & \textcolor{violet}{\textbf{27}} & $\Upsilon(1S) \rm \ B_{c}^{\pm}(1 \prescript{3}{}{D}_{3})$
\\
\cline{3-9}
 & & \multirow{7}{*}{2P} & 1 & $0^{-}$ & 16,631 & 16,098 & 533 & $\rm \eta_{b}(1S) \ B_{c}^{\pm}(1 \prescript{3}{}{P}_{0})$
\\
\cline{4-9}
 & & & 0 & \multirow{3}{*}{$1^{-}$} & 16,629 & \multirow{3}{*}{16,142} & 487 & \multirow{3}{*}{$\rm \eta_{b}(1S) \ B_{c}^{\pm}(1 \prescript{ }{}{P}_{1})$}
\\
 & & & 1 & & 16,630 & & 488 & 
\\
 & & & 2 & & 16,633 & & 491 & 
\\
\cline{4-9}
 & & & 1 & \multirow{2}{*}{$2^{-}$} & 16,630 & \multirow{2}{*}{16,160} & 470 & \multirow{2}{*}{$\rm \eta_{b}(1S) \ B_{c}^{\pm}(1 \prescript{3}{}{P}_{2})$}
\\
 & & & 2 & & 16,632 & & 472 & 
\\
\cline{4-9}
 & & & 2 & $3^{-}$ & 16,631 & 16,221 & 410 & $\Upsilon(1S) \rm \ B_{c}^{\pm}(1 \prescript{3}{}{P}_{2})$
\\
\cline{3-9}
 & & \multirow{3}{*}{3S} & 0 & $0^{+}$ & 16,765 & 15,673 & 1092 & $\rm \boldsymbol \eta_{b}\mathbf{(1S) \ B_{c}^{\pm}(1 \prescript{1}{}{S}_{0})}$
\\
\cline{4-9}
 & & & 1 & $1^{+}$ & 16,765 & 15,732 & 1033 & $\rm \eta_{b}(1S) \ B_{c}^{\pm}(1 \prescript{3}{}{S}_{1})$
\\
\cline{4-9}
 & & & 2 & $2^{+}$ & 16,764 & 15,793 & 881 & $\Upsilon(1S) \rm \ B_{c}^{\pm}(1 \prescript{3}{}{S}_{1})$
\\
\cline{2-9}
 & \multirow{12}{*}{$\rm S\overline A$, $\rm A\overline S$} & 1S & \multirow{12}{*}{1} & $1^{+}$ & 16,099 & 15,732 & 367 & $\rm \eta_{b}(1S) \ B_{c}^{\pm}(1 \prescript{3}{}{S}_{1})$
\\
\cline{3-3}
\cline{5-9}
 & & \multirow{3}{*}{1P} & & $0^{-}$ & 16,320 & 16,098 & 222 & $\rm \eta_{b}(1S) \ B_{c}^{\pm}(1 \prescript{3}{}{P}_{0})$
\\
\cline{5-9}
 & & & & $1^{-}$ & 16,321 & 16,142 & 179 & $\rm \eta_{b}(1S) \ B_{c}^{\pm}(1 \prescript{ }{}{P}_{1})$
\\
\cline{5-9}
 & & & & $2^{-}$ & 16,322 & 16,160 & 162 & $\rm \eta_{b}(1S) \ B_{c}^{\pm}(1 \prescript{3}{}{P}_{2})$
\\
\cline{3-3}
\cline{5-9}
 & & 2S & & $1^{+}$ & 16,474 & 15,732 & 742 & $\rm \eta_{b}(1S) \ B_{c}^{\pm}(1 \prescript{3}{}{S}_{1})$
\\
\cline{3-3}
\cline{5-9}
 & & \multirow{3}{*}{1D} & & $1^{+}$ & 16,507 & 15,732 & 775 & $\rm \eta_{b}(1S) \ B_{c}^{\pm}(1 \prescript{3}{}{S}_{1})$
\\
\cline{5-9}
 & & & & $2^{+}$ & 16,508 & 15,793 & 715 & $\Upsilon(1S) \rm \ B_{c}^{\pm}(1 \prescript{3}{}{S}_{1})$
\\
\cline{5-9}
 & & & & \textcolor{violet}{$\mathbf{3^{+}}$} & \textcolor{violet}{\textbf{16,509}} & 16,428 & \textcolor{violet}{\textbf{81}} & $\rm \eta_{b}(1S) \ B_{c}^{\pm}(1 \prescript{3}{}{D}_{3})$
\\
\cline{3-3}
\cline{5-9}
 & & \multirow{3}{*}{2P} & & $0^{-}$ & 16,624 & 16,098 & 526 & $\rm \eta_{b}(1S) \ B_{c}^{\pm}(1 \prescript{3}{}{P}_{0})$
\\
\cline{5-9}
 & & & & $1^{-}$ & 16,624 & 16,142 & 482 & $\rm \eta_{b}(1S) \ B_{c}^{\pm}(1 \prescript{ }{}{P}_{1})$
\\
\cline{5-9}
 & & & & $2^{-}$ & 16,624 & 16,160 & 464 & $\rm \eta_{b}(1S) \ B_{c}^{\pm}(1 \prescript{3}{}{P}_{2})$
\\
\cline{3-3}
\cline{5-9}
 & & 3S & & $1^{+}$ & 16,758 & 15,732 & 1026 & $\rm \eta_{b}(1S) \ B_{c}^{\pm}(1 \prescript{3}{}{S}_{1})$
\end{tabular}
\end{ruledtabular}
\end{table*}

\par
From the Tables~\ref{Tab:Thrcccb}-\ref{Tab:Thrbbbc}, a number of conclusions can be drawn. In particular, we can distinguish three types of tetraquark states: those lying significantly above, slightly above and below the fall-apart strong decay thresholds.
\begin{itemize}

	\item For all of the considered ground and excited states of asymmetric fully heavy tetraquarks, with the exception of two higher excitations (they will be discussed separately), there is at least one heavy meson pair with a total mass less than the tetraquark mass ($\Delta_{\rm max} > 0$). Therefore, for almost all of the considered states of tetraquarks there is a possibility of such fall-apart strong decay. Moreover, for the vast majority of considered states of tetraquarks the value of the $\Delta_{max}$ significantly exceeds $100$ MeV and for many of them it even exceeds $500$ MeV. As we have discussed in the beginning of this section, experimentally such states could manifest themselves as wide resonances. However, this statement is true only for the ground states of tetraquarks. For the excited states the additional restrictions appear. In particular, these decays can be suppressed either by the centrifugal barrier between the quark and the antiquark (for the orbital excitations), or by the zeros of the wave function (for the radial excitations), or both, and therefore they still can end up being narrow resonances.

	\item Nonetheless, there is a number of states for which $\Delta_{max} \leq 100$ MeV. Although, such states lie above, they are already close to the threshold of the fall-apart strong decay into a pair of mesons. Thus, in Tables~\ref{Tab:Thrcccb}-\ref{Tab:Thrbbbc} they are highlighted in purple as promising to be relatively narrow. 
	
	\item Finally, there are two states (highlighted in red in Tables~\ref{Tab:Thrcccb}-\ref{Tab:Thrbbbc}) which lie below any of the possible heavy meson pair thresholds with their $\Delta_{max} < 0$:
	\begin{widetext}
	\begin{gather}
	{\rm cc\overline c\overline b, bc\overline c\overline c} \qquad {\rm A\overline A} \qquad {\rm 1D} \qquad J^{P}=4^{+} \qquad M=10,114 \ {\rm MeV}, \label{States:cccb}
	\\
	{\rm cc\overline b\overline b, bb\overline c\overline c} \qquad {\rm A\overline A} \qquad {\rm 1D} \qquad J^{P}=4^{+} \qquad M=13,329 \ {\rm MeV}. \label{States:ccbb}
	\end{gather}
	\end{widetext}
	\noindent
	Note that our calculations have a theoretical error: the uncertainties of our calculations come from the quark model and diquark--antidiquark approximation. We roughly estimate them from our previous experience to be about $20-50$ MeV. Thus, if $\Delta_{\rm max}$ is slightly negative, it does not mean for sure that the state cannot fall-apart into a meson pair. However, if the value of $\Delta_{\rm max}$ is sufficiently negative than the state definitely cannot decay via the strong fall-apart strong decay processes into two heavy mesons. In this case the main channels will be a strong decay due to the heavy quark--antiquark annihilation into gluons with their subsequent hadronization into the lighter hadrons or radiative decays, if allowed. As a result, this state could be a narrow state that could be observed experimentally in other decay channels, either to hadrons made up of lighter quarks and antiquarks or two heavy mesons and a photon.
	
\end{itemize}
\noindent
In the Table~\ref{Tab:ThrProb} we additionally present states of the asymmetric fully heavy tetraquarks which lie below or slightly above ($\Delta_{\rm max} \lesssim 100$ MeV) the meson--meson fall-apart strong decay thresholds. Such states have the most chances to be observed in experiments, as it was discussed above. Also in Tables~\ref{Tab:Thrcccb}-\ref{Tab:Thrbbbc} we highlight in bold all lowest thresholds containing the $B_{c}^{\pm}(1 \prescript{1}{}{S}_{0})$-meson. The reason is that currently only two charmed--bottom mesons were observed~\cite{PDG2022} --- $B_{c}^{\pm}(1 \prescript{1}{}{S}_{0})$ and $B_{c}^{\pm}(2 \prescript{1}{}{S}_{0})$, both with total momentum--parity $J^{P}=0^{-}$.

\begin{table*}
\caption{Excited states of the asymmetric ($\rm cc\overline c\overline b$, $\rm bc\overline c\overline c$, $\rm cc\overline b\overline b$, $\rm bb\overline c\overline c$, $\rm bb\overline b\overline c$, $\rm cb\overline b\overline b$) fully heavy tetraquarks which lie slightly above or below the meson--meson fall-apart strong decay thresholds. All notations are explained in the caption of Table~\ref{Tab:Thrcccb}.\label{Tab:ThrProb}}
\begin{ruledtabular}
\begin{tabular}{ccccccccc}
$\rm QQ'\overline Q\overline Q'$ & $\rm d\overline d'$ & $n\rm L$ & $S$ & $J^{P}$ & $M_{\rm QQ'\overline Q\overline Q'}$ & $M_{\rm thr.}$ & $\Delta_{\rm max}$ & Meson pair
\\
\hline
\multirow{5}{*}{\makecell{$\rm cc\overline c\overline b$, \\ $\rm bc\overline c\overline c$}} & \multirow{4}{*}{$\rm A\overline A$} & 1P & 2 & $3^{-}$ & 9,881 & 9,858 & 23 & $\rm J/\psi(1S) \ B_{c}^{\pm}(1 \prescript{3}{}{P}_{2})$
\\
\cline{3-9}
 & & \multirow{3}{*}{1D} & 1 & \multirow{2}{*}{$3^{+}$} & 10,111 & \multirow{2}{*}{10,013} & 98 & \multirow{2}{*}{$\rm \eta_{c}(1S) \ B_{c}^{\pm}(1 \prescript{3}{}{D}_{3})$}
\\
 & & & 2 & & 10,116 & & 103 & 
\\
\cline{4-9}
 & & & 2 & $4^{+}$ & 10,114 & 10,126 & -12 & $\rm J/\psi(1S) \ B_{c}^{\pm}(1 \prescript{3}{}{D}_{3})$
\\
\cline{2-9}
 & $\rm S\overline A$, $\rm A\overline S$ & 1D & 1 & $3^{+}$ & 10,105 & 10,013 & 92 & $\rm \eta_{c}(1S) \ B_{c}^{\pm}(1 \prescript{3}{}{D}_{3})$
\\
\hline
\multirow{9}{*}{\makecell{$\rm cc\overline b\overline b$, \\ $\rm bb\overline c\overline c$}} & \multirow{9}{*}{$\rm A\overline A$} & \multirow{6}{*}{1P} & 0 & \multirow{3}{*}{$1^{-}$} & 13,103 & \multirow{3}{*}{13,017} & 86 & \multirow{3}{*}{$\rm B_{c}^{\pm}(1 \prescript{1}{}{S}_{0}) \ B_{c}^{\pm}(1 \prescript{ }{}{P}_{1})$}
\\
 & & & 1 & & 13,108 & & 91 &
\\
 & & & 2 & & 13,111 & & 94 &
\\
\cline{4-9}
 & & & 1 & \multirow{2}{*}{$2^{-}$} & 13,106 & \multirow{2}{*}{13,035} & 71 & \multirow{2}{*}{$\rm B_{c}^{\pm}(1 \prescript{1}{}{S}_{0}) \ B_{c}^{\pm}(1 \prescript{3}{}{P}_{2})$}
\\
 & & & 2 & & 13,112 & & 77 &
\\
\cline{4-9}
 & & & 2 & $3^{-}$ & 13,110 & 13,094 & 16 & $\rm B_{c}^{\pm}(1 \prescript{3}{}{S}_{1}) \ B_{c}^{\pm}(1 \prescript{3}{}{P}_{2})$
\\
\cline{3-9}
 & & \multirow{3}{*}{1D} & 1 & \multirow{2}{*}{$3^{+}$} & 13,327 & \multirow{2}{*}{13,303} & 24 & \multirow{2}{*}{$\rm B_{c}^{\pm}(1 \prescript{1}{}{S}_{0}) \ B_{c}^{\pm}(1 \prescript{3}{}{D}_{3})$}
\\
 & & & 2 & & 13,332 & & 29 & 
\\
\cline{4-9}
 & & & 2 & $4^{+}$ & 13,329 & 13,362 & -33 & $\rm B_{c}^{\pm}(1 \prescript{3}{}{S}_{1}) \ B_{c}^{\pm}(1 \prescript{3}{}{D}_{3})$
\\
\hline
\multirow{4}{*}{\makecell{$\rm bb\overline b\overline c$, \\ $\rm cb\overline b\overline b$}} & \multirow{3}{*}{$\rm A\overline A$} & \multirow{3}{*}{1D} & 1 & \multirow{2}{*}{$3^{+}$} & 16,515 & \multirow{2}{*}{16,428} & 87 & \multirow{2}{*}{$\rm \eta_{b}(1S) \ B_{c}^{\pm}(1 \prescript{3}{}{D}_{3})$}
\\
 & & & 2 & & 16,516 & & 88 & 
\\
\cline{4-9}
 & & & 2 & $4^{+}$ & 16,516 & 16,489 & 27 & $\Upsilon(1S) \rm \ B_{c}^{\pm}(1 \prescript{3}{}{D}_{3})$
\\
\cline{2-9}
 & $\rm S\overline A$, $\rm A\overline S$ & 1D & 1 & $3^{+}$ & 16,509 & 16,428 & 81 & $\rm \eta_{b}(1S) \ B_{c}^{\pm}(1 \prescript{3}{}{D}_{3})$
\end{tabular}
\end{ruledtabular}
\end{table*}

\par
Let us discuss the recent experimental progress. As we have already mentioned in Sec.~\ref{Sec:Model} the symmetric fully heavy flavor compositions are easier to obtain. Moreover, the first experimental data for the symmetric states of the fully charmed tetraquark (X(6900) and a few broad peaks in the 6.4-7.4 GeV mass region) were recently reported by the LHCb~\cite{cccc:LHCb:2020}, ATLAS~\cite{cccc:ATLAS:2023} and CMS~\cite{cccc:CMS:2023} Collaborations. All experiments indicate the existence of at least three structures which can be denoted as X(6600), X(6900) and X(7200). In Table~\ref{Tab:ExpSym} we compare these data with our previous predictions from Ref.~\cite{Savch2022Sym} and propose possible quantum numbers $J^{PC}$ for the observed states.

\begin{table*}
\caption{Exotic X states observed and hinted by the LHCb~\cite{cccc:LHCb:2020}, ATLAS~\cite{cccc:ATLAS:2023} and CMS~\cite{cccc:CMS:2023} Collaborations in di-$J/\psi$ and $J/\psi \, \psi(2S)$ invariant mass spectra and our candidates~\cite{Savch2022Sym}. $n\rm L$ denotes the overall excitation. $S$ is the total spin of the diquark--antidiquark system. All masses $M$ and total widths $\Gamma$ are given in MeV.\label{Tab:ExpSym}}
\begin{ruledtabular}
\begin{tabular}{rrccccccc}
\multicolumn{2}{c}{\multirow{2}{*}{\textbf{Collaboration}}} & \multirow{2}{*}{\textbf{Resonance}} & \multirow{2}{*}{$\mathbf{M}$} & \multirow{2}{*}{$\boldsymbol \Gamma$} & \multicolumn{4}{c}{\textbf{Our candidates}}
\\
\cline{6-9}
 & & & & & $n\rm L$ & $S$ & $\mathbf{J^{PC}}$ & $\mathbf{M}$
\\
\hline
\multicolumn{2}{c}{LHCb} & \multirow{6}{*}{X(6600)} & 6,400 $\divisionsymbol$ 6,600 & & \multirow{3}{*}{1S} & \multirow{3}{*}{2} & \multirow{3}{*}{$2^{++}$} & \multirow{3}{*}{6,367}
\\
\cline{1-2}
\cline{4-5}
\multirow{3}{*}{ATLAS} & {\tiny $\rm m_{0}$, model A} & & $6,410 \pm 80^{+80}_{-30}$ & $590 \pm 350^{+120}_{-200}$ & & & & 
\\
\cline{4-5}
 & {\tiny $\rm m_{0}$, model B} & & $6,650 \pm 20^{+30}_{-20} $ & $440 \pm 50^{+60}_{-50}$ & & & & 
\\
\cline{4-5}
 & {\tiny $\rm m_{1}$, model A} & & $6,630 \pm 50^{+80}_{-10} $ & $350 \pm 110^{+110}_{-40}$ & \multirow{3}{*}{2S} & \multirow{3}{*}{0} & \multirow{3}{*}{$0^{++}$} & \multirow{3}{*}{6,782} 
\\
\cline{1-2}
\cline{4-5}
\multirow{2}{*}{CMS} & {\tiny \makecell{$\rm BW_{1}$, \\ no interference}} & & $6,552 \pm 10 \pm 12$ & $124^{+32}_{-26} \pm 33$ & & & & 
\\
\cline{4-5}
 & {\tiny \makecell{$\rm BW_{1}$, \\ interference}} & & $6,638^{+43+16}_{-38-31}$ & $440^{+230+110}_{-200-240}$ & & & & 
\\
\hline
\multirow{2}{*}{LHCb} & {\tiny \makecell{NRSPS, \\ no interference}} & \multirow{7}{*}{X(6900)} & $6,905 \pm 11 \pm 7$ & $80 \pm 19 \pm 33$ & & & & 
\\
\cline{4-5}
 & {\tiny \makecell{NRSPS, \\ interference}} & & $6,886 \pm 11 \pm 11$ & $168 \pm 33 \pm 69$ & 2S & 2 & $2^{++}$ & 6,868 
\\
\cline{1-2}
\cline{4-5}
\multirow{3}{*}{ATLAS} & {\tiny $\rm m_{2}$, model A} & & $6,860 \pm 30^{+10}_{-20}$ & $110 \pm 50^{+20}_{-10}$ & \multirow{4}{*}{1D} & 0 & $2^{++}$ & 6,921 
\\
\cline{4-5}
 & {\tiny $\rm m_{2}$, model B} & & $6,910 \pm 10 \pm 10 $ & $150 \pm 30 \pm 10$ & & 2 & $0^{++}$ & 6,899 
\\
\cline{4-5}
 & {\tiny $\rm m_{3}$, model $\beta$} & & $6,960 \pm 50 \pm 30$ & $510 \pm 170^{+110}_{-100}$ & & 2 & $1^{++}$ & 6,904 
\\
\cline{1-2}
\cline{4-5}
\multirow{2}{*}{CMS} & {\tiny \makecell{$\rm BW_{2}$, \\ no interference}} & & $6,927 \pm 9 \pm 4$ & $122^{+24}_{-21} \pm 18$ & & 2 & $2^{++}$ & 6,915 
\\
\cline{4-5}
 & {\tiny \makecell{$\rm BW_{2}$, \\ interference}} & & $6,847^{+44+48}_{-28-20}$ & $191^{+66+25}_{-49-17}$ & & & & 
\\
\hline
\multicolumn{2}{c}{LHCb} & \multirow{4}{*}{X(7200)} & 7,200 $\divisionsymbol$ 7,400 & & \multirow{4}{*}{3S} & \multirow{2}{*}{0} & \multirow{2}{*}{$0^{++}$} & \multirow{2}{*}{7,259}  
\\
\cline{1-2}
\cline{4-5}
ATLAS & {\tiny $\rm m_{3}$, model $\alpha$} & & $7,220 \pm 30^{+10}_{-30}$ & $90 \pm 60^{+60}_{-30}$ & & & & 
\\
\cline{1-2}
\cline{4-5}
\multirow{2}{*}{CMS} & {\tiny \makecell{$\rm BW_{3}$, \\ no interference}} & & $7,287^{+20}_{-18} \pm 5$ & $95^{+59}_{-40} \pm 19$ & & \multirow{2}{*}{2} & \multirow{2}{*}{$2^{++}$} & \multirow{2}{*}{7,333} 
\\
\cline{4-5}
 & {\tiny \makecell{$\rm BW_{3}$, \\ interference}} & & $7,134^{+48+41}_{-25-15}$ & $97^{+40+29}_{-29-26}$ & & & & 
\end{tabular}
\end{ruledtabular}
\end{table*}

\hfill 
\par
Unfortunately, no experimental data on asymmetric fully heavy tetraquark states are available yet. Nonetheless, our calculations~\eqref{States:cccb},~\eqref{States:ccbb} show that there is a fair possibility to observe at least some of the higher excitations in D-wave in the future.

\hfill 
\section{Theoretical Predictions}
\label{Sec:Theor}

\par
It is important to compare the predictions for the masses of the asymmetric fully heavy tetraquarks in different theoretical approaches in order to identify the most promising candidates for the experimental searches and their mass ranges. We present such comparison in the Tables~\ref{Tab:Comp.cccb.AA.1S}-\ref{Tab:Comp.bbbc.SA.1S}.
\par
Approaches in the considered papers can be divided into the following general groups.
\begin{itemize}

	\item Various quark models \cite{QQQQ.ground:all.2018.1, QQQQ.ground:all.2018.2, QQQQ.ground:cbcb;ccbb.2019.1, QQQQ:ccbb.2019.2, QQQQ.ground:all.2019.3, QQQQ.ground;exc:cccc;bbbb;ccbb.2019.4, qqQQ.ground:sbsb;ssbb;qqbb.QQQQ.ground;exc:cccc;cbcb;bbbb;ccbb.2020.1, QQQQ.ground:all.2020.2, QQQQ.ground:all.exc:ccbb.2021.1, QQQQ.ground:all.2021.3, QQQQ.ground:all.2021.4, QQQQ.ground:cccb;ccbb;bbbc.2022.2, QQQQ.ground:all.2022.3, QQQQ.ground:all.2023.1};
	
	\item QCD sum rules \cite{QQQQ.ground:ccbb.2021.2, QQQQ.ground:cccc;cbcb;bbbb;ccbb.2022.1};
	
	\item Lattice QCD \cite{qqQQ:udbb;ssbb.QQQQ:ccbb.2015.1}.
	
\end{itemize}
\noindent
We can further distinguish several rather popular approaches which are used commonly. For them we introduce the following abbreviations which are utilized in the Tables~\ref{Tab:Comp.cccb.AA.1S}-\ref{Tab:Comp.bbbc.SA.1S}.
\begin{itemize}
	
	\item Interpretations of the 4-quark state structure
	\begin{itemize}
	
		\item DA --- diquark--antidiquark picture,
		
		\item MM --- meson--meson model,
		
		\item mix --- mixture of the models above.
		
	\end{itemize}
	
	\item Mathematical models
	\begin{itemize}
	
		\item CM --- chromomagnetic model,
	
		\item MCFTM --- multiquark color flux-tube model,
	
		\item DDM --- dynamical diquark model,
	
		\item RDM --- relativized diquark model,
	
		\item NChQM --- nonrelativistic chiral quark model,
	
		\item NCoQM --- nonrelativistic constituent quark model.
		
	\end{itemize}
	
	\item Computational methods
	\begin{itemize}
	
		\item DMCM --- diffusion Monte Carlo method,
	
		\item QCDSR --- QCD sum rules,
	
		\item VA --- variational approach.
		
	\end{itemize}
	
\end{itemize}
\noindent
Note, that in Ref.~\cite{QQQQ.ground:all.2022.3} the authors use for the spin-dependent interaction either the one-gluon exchange (OGE) or instanton-induced interaction (INS). Additionally, in Tables~\ref{Tab:Comp.cccb.AA.1S}-\ref{Tab:Comp.bbbc.SA.1S} we highlight results obtained in the diquark--antidiquark picture which is the same as ours interpretation of the four-quark state structure.
\par
A few remarks on the results presented in the Tables~\ref{Tab:Comp.cccb.AA.1S}-\ref{Tab:Comp.bbbc.SA.1S} are necessary.
\begin{itemize}
	
	\item \textcolor{black}{The study of asymmetric $cc\overline b\overline b$ tetraquark using the similar to ours assumptions was performed in the relativized diquark model in Ref.~\cite{qqQQ.ground:sbsb;ssbb;qqbb.QQQQ.ground;exc:cccc;cbcb;bbbb;ccbb.2020.1}. The color-($\overline 3 \times 3$) diquarks and antidiquarks were considered with the excitations occurring only between them. The main difference from our approach consists in the construction of the relativistic interaction potential. We systematically include all spin-dependent and spin-independent contributions and take into account the finite diquark size while the authors of Ref.~\cite{qqQQ.ground:sbsb;ssbb;qqbb.QQQQ.ground;exc:cccc;cbcb;bbbb;ccbb.2020.1} take only spin-dependent relativistic corrections to the one-gluon exchange potential and consider diquarks to be point-like. Note that the spin--orbit interaction term is missing in their potential. Moreover, while we solve numerically the quasipotential Eq.~\eqref{Eq:Schr} with the quasipotential \eqref{Eq:V} in Ref.~\cite{qqQQ.ground:sbsb;ssbb;qqbb.QQQQ.ground;exc:cccc;cbcb;bbbb;ccbb.2020.1} the variational procedure with the harmonic oscillator trial wave functions is employed.}
	
	\item Most of the researches within the diquark–-antidiquark picture consider both possible color structures (and sometimes also the mixture of them): diquark--antidiquark in the ($\overline 3 \times 3$) and ($6 \times \overline 6$) color configurations. In Secs.~\ref{Sec:Model},~\ref{Sec:Res} we have already thoroughly discussed why we consider the latter case inappropriate for our problem (to put it in a few words, the internal quark--quark interaction within the color-sextet diquark is repulsive, thus, it cannot be a bound state \textcolor{black}{if diquarks interact as a whole}). Additionally, the color-sextet diquarks and -antisextet antidiquarks can simultaneously have $S_{\rm d}=0$, making possible the color-($6 \times \overline 6$) tetraquarks of scalar--scalar spin configuration. On the contrary, the fully heavy color-antitriplet diquarks and -triplet antidiquarks can produce only axialvector--axialvector and/or scalar--axialvector tetraquarks. It is important, that in the Tables~\ref{Tab:Comp.cccb.AA.1S}-\ref{Tab:Comp.bbbc.SA.1S} we present the results within the specific diquark--antidiquark spin configurations. Hence, in these Tables we omit the results for the color-($6 \times \overline 6$) scalar--scalar tetraquarks which are impossible within our framework. Note, that in literature there is no consensus in how the color configuration should affect the mass spectra of fully heavy tetraquarks. \textcolor{black}{For example in Ref.~\cite{QQQQ.ground:all.2019.3} the masses of the ground states of the fully heavy tetraquarks were calculated in the potential quark model with linear confinement, Coulomb and spin--spin interactions. Both color ($\overline 3 \times 3$) and ($6 \times \overline 6$) color components as well as their mixture were considered in a four-body fashion using Gaussian functions.} In Refs.~\cite{QQQQ.ground:all.2018.2, QQQQ.ground:all.2019.3, QQQQ.ground:all.2022.3} the calculated masses of the sextet--antisextet color configurations lie approximately $5-200$ MeV higher than their antitriplet--triplet counterparts. On the contrary, in Refs.~\cite{QQQQ.ground;exc:cccc;bbbb;ccbb.2019.4, QQQQ.ground:all.2021.3, QQQQ.ground:all.2021.4, QQQQ.ground:all.2023.1}, the masses of the sextet--antisextet color configurations are approximately $5-200$ MeV lower than their antitriplet--triplet counterparts.
	
	\item In papers~\cite{QQQQ.ground:ccbb.2021.2, QQQQ.ground:cccc;cbcb;bbbb;ccbb.2022.1} the excited P-wave diquarks (pseudoscalar, vector and pseudotensor with $J^{P}=0^{-}, \ 1^{-}, \ 2^{-}$) were also considered. This leads to the emergence of more possible diquark--antidiquark spin configurations, not available within our framework. Also, recalling our discussion from Sec.~\ref{Sec:Res}, we consider the tetraquarks containing excited diquarks to be less stable due to the enhanced diquark--antiquark overlap leading to faster fall-apart strong decay. Therefore, in the Tables~\ref{Tab:Comp.cccb.AA.1S}-\ref{Tab:Comp.bbbc.SA.1S} we present results only for the axialvector--axialvector and scalar--axialvector tetraquarks which are built from the ground S-wave diquarks.
	
	\item Authors considering two constituent color structures, i.e. ($\overline 3 \times 3$) and ($6 \times \overline 6$) for the diquark–-antidiquark picture and ($1 \times 1$) and ($8 \times 8$) for the meson--meson model, usually assume them to be the pure states and/or allow their mixture (or coupling) with varying percentage of each component. If within one paper there are several results we explicitly identify the color state for which the result was obtained.
	
\end{itemize}

\begin{table*}
\caption{Comparison of different theoretical predictions for the masses of the triple charmed and bottom fully heavy tetraquark ($\rm cc\overline c\overline b$, $\rm bc\overline c\overline c$) with the axialvector--axialvector spin configuration in the ground state. $\rm d$ and $\rm \overline d'$ are the axialvector (A) or scalar (S) diquark and antidiquark, respectively. $n\rm L$ denotes the overall excitation. $S$ is the total spin of the diquark--antidiquark system. $J^{P}$ is the total momentum-parity of the tetraquark. $M_{\rm thr.}$ is the corresponding meson–-meson threshold~\cite{PDG2022, Regge2011meson}. All masses are given in MeV. Results are sorted chronologically, oldest predictions first.\label{Tab:Comp.cccb.AA.1S}}
\begin{ruledtabular}
\begin{tabular}{cccc}
$\rm d\overline d'$ & \multicolumn{3}{c}{$\rm A\overline A$}
\\
\hline
$n\rm L$ & \multicolumn{3}{c}{1S}
\\
\hline
$S$ & 0 & 1 & 2
\\
\hline
$J^{P}$ & $0^{+}$ & $1^{+}$ & $2^{+}$
\\
\hline
$M_{\rm thr.}$ & 9,258 & 9,317 & 9,430
\\
\hline
Our & 9,606 & 9,611 & 9,620
\\
\hline 
\cite{QQQQ.ground:all.2018.1}{\scriptsize (VA DA)} & \multicolumn{3}{c}{$\lesssim$9,390}
\\
\hline
\cite{QQQQ.ground:all.2018.2}{\scriptsize (CM)} & \linebreakvertS{9,313 {\scriptsize [MM threshold]} \\ 10,144 {\scriptsize [DA, ($\overline 3 \times 3$)-$\rm A\overline A$ mixed with ($6 \times \overline 6$)-$\rm S\overline S$]}} & \linebreakvertS{9,400 {\scriptsize [MM threshold]} \\ 10,231 {\scriptsize [DA, ($\overline 3 \times 3$)-$\rm A\overline A$ mixed with -$\rm A\overline S$ and ($6 \times \overline 6$)-$\rm S\overline A$]}} & \linebreakvertS{9,442 {\scriptsize [MM threshold]} \\ 10,273 {\scriptsize [DA]}}
\\
\hline
\cite{QQQQ.ground:all.2019.3}{\scriptsize (NCoQM DA)} & 9,740 & 9,749 & 9,768
\\
\hline
\cite{QQQQ.ground:all.2020.2}{\scriptsize (DMCM DA)} & \linebreakvertS{9,615 {\scriptsize [($\overline 3 \times 3$)-$\rm A\overline A$ coupled with ($6 \times \overline 6$)-$\rm S\overline S$]}} & \linebreakvertS{9,610 {\scriptsize [($\overline 3 \times 3$)-$\rm A\overline A$ coupled with -$\rm S\overline A$ and ($6 \times \overline 6$)-$\rm A\overline S$]}} & 9,719
\\
\hline
\cite{QQQQ.ground:all.exc:ccbb.2021.1}{\scriptsize (DA)} & \linebreakvertS{9,314 {\scriptsize [CM, ($\overline 3 \times 3$)-$\rm A\overline A$ coupled with ($6 \times \overline 6$)-$\rm S\overline S$]} \\ 9,670 {\scriptsize [MCFTM, ($\overline 3 \times 3$)-$\rm A\overline A$ coupled with ($6 \times \overline 6$)-$\rm S\overline S$]} \\ 9,705 {\scriptsize [MCFTM, ($\overline 3 \times 3$)-$\rm A\overline A$]} \\ 9,753 {\scriptsize [NCoQM, ($\overline 3 \times 3$)-$\rm A\overline A$ coupled with ($6 \times \overline 6$)-$\rm S\overline S$]} \\ 9,813 {\scriptsize [NCoQM, ($\overline 3 \times 3$)-$\rm A\overline A$]}} & \linebreakvertS{9,343 {\scriptsize [CM, ($\overline 3 \times 3$)-$\rm A\overline A$ coupled with -$\rm A\overline S$ and ($6 \times \overline 6$)-$\rm S\overline A$]} \\ 9,683 {\scriptsize [MCFTM, ($\overline 3 \times 3$)-$\rm A\overline A$ coupled with -$\rm A\overline S$ and ($6 \times \overline 6$)-$\rm S\overline A$]} \\ 9,705 {\scriptsize [MCFTM, ($\overline 3 \times 3$)-$\rm A\overline A$ coupled with -$\rm A\overline S$]} \\ 9,766 {\scriptsize [NCoQM, ($\overline 3 \times 3$)-$\rm A\overline A$ coupled with -$\rm A\overline S$ and ($6 \times \overline 6$)-$\rm S\overline A$]} \\ 9,808 {\scriptsize [NCoQM, ($\overline 3 \times 3$)-$\rm A\overline A$ coupled with -$\rm A\overline S$]}} & \linebreakvertS{9,442 {\scriptsize [CM]} \\ 9,732 {\scriptsize [MCFTM]} \\ 9,839 {\scriptsize [NCoQM]}}
\\
\hline
\cite{QQQQ.ground:all.2021.3}{\scriptsize (CM DA)} & \linebreakvertS{9,505.9 {\scriptsize [($\overline 3 \times 3$)-$\rm A\overline A$ mixed with ($6 \times \overline 6$)-$\rm S\overline S$]}} & \linebreakvertS{9,484.3 {\scriptsize [($\overline 3 \times 3$)-$\rm A\overline A$ mixed with -$\rm A\overline S$ and ($6 \times \overline 6$)-$\rm S\overline A$]}} & 9,525.9
\\
\hline
\cite{QQQQ.ground:all.2021.4}{\scriptsize (NCoQM)} & \linebreakvertS{9,243 {\scriptsize [MM, ($1 \times 1$) coupled]} \\ 9,559 {\scriptsize [MM, ($8 \times 8$) coupled]} \\ 9,620 {\scriptsize [DA, ($\overline 3 \times 3$)-$\rm A\overline A$ coupled with ($6 \times \overline 6$)-$\rm S\overline S$]} \\ 9,673 {\scriptsize [DA, ($\overline 3 \times 3$)-$\rm A\overline A$]} \\ 9,310, 9,551, 9,599, 9,621 {\scriptsize [other coupled spatial configurations]}} & \linebreakvertS{9,317 {\scriptsize [MM, ($1 \times 1$) coupled]} \\ 9,587 {\scriptsize [MM, ($8 \times 8$) coupled]} \\ 9,638 {\scriptsize [DA, ($\overline 3 \times 3$)-$\rm A\overline A$ coupled with -$\rm A\overline S$ and ($6 \times \overline 6$)-$\rm S\overline A$]} \\ 9,687 {\scriptsize [DA, ($\overline 3 \times 3$)-$\rm A\overline A$]} \\ 9,546, 9,568, 9,619, 9,640 {\scriptsize [other coupled spatial configurations]}} & \linebreakvertS{9,451 {\scriptsize [MM, ($1 \times 1$)]} \\ 9,650 {\scriptsize [MM, ($8 \times 8$)]} \\ 9,714 {\scriptsize [DA, ($\overline 3 \times 3$)]} \\ 9,593, 9,688, 9,704, 9,717 {\scriptsize [other coupled spatial configurations]}}
\\
\hline
\cite{QQQQ.ground:cccb;ccbb;bbbc.2022.2}{\scriptsize (DDM DA)} & \linebreakvertS{9,560 {\scriptsize [set $\RNumb{2}$]} \\ 9,579 {\scriptsize [set $\RNumb{1}$]}} & \linebreakvertS{9,571 {\scriptsize [set $\RNumb{2}$]} \\ 9,590 {\scriptsize [set $\RNumb{1}$]}} & \linebreakvertS{9,594 {\scriptsize [set $\RNumb{2}$]} \\ 9,613 {\scriptsize [set $\RNumb{1}$]}}
\\
\hline
\cite{QQQQ.ground:all.2022.3}{\scriptsize (NCoQM DA)} & \linebreakvertS{9,615 {\scriptsize [INS, ($\overline 3 \times 3$)-$\rm A\overline A$ mixed with ($6 \times \overline 6$)-$\rm S\overline S$]} \\ 9,665 {\scriptsize [OGE, ($\overline 3 \times 3$)-$\rm A\overline A$ mixed with ($6 \times \overline 6$)-$\rm S\overline S$]}} & \linebreakvertS{9,646 {\scriptsize [INS, ($\overline 3 \times 3$)-$\rm A\overline A$ mixed with -$\rm A\overline S$ and ($6 \times \overline 6$)-$\rm S\overline A$]} \\ 9,699 {\scriptsize [OGE, ($\overline 3 \times 3$)-$\rm A\overline A$ mixed with -$\rm A\overline S$ and ($6 \times \overline 6$)-$\rm S\overline A$]}} & \linebreakvertS{9,645 {\scriptsize [INS]} \\ 9,713 {\scriptsize [OGE]}}
\\
\hline
\cite{QQQQ.ground:all.2023.1}{\scriptsize (NCoQM DA)} & \linebreakvertS{9,766 {\scriptsize [($\overline 3 \times 3$)-$\rm A\overline A$ mixed with ($6 \times \overline 6$)-$\rm S\overline S$]}} & \linebreakvertS{9,706 {\scriptsize [($\overline 3 \times 3$)-$\rm A\overline A$ mixed with -$\rm A\overline S$ and ($6 \times \overline 6$)-$\rm S\overline A$]}} & 9,731
\end{tabular}
\end{ruledtabular}
\end{table*}

\begin{table*}
\caption{Same as in Table~\ref{Tab:Comp.cccb.AA.1S}, but for the scalar--axialvector spin configuration in the ground state. Notations are explained in the caption of Table~\ref{Tab:Comp.cccb.AA.1S}.\label{Tab:Comp.cccb.SA.1S}}
\begin{ruledtabular}
\begin{tabular}{cc}
$\rm d\overline d'$ & $\rm S\overline A$, $\rm A\overline S$
\\
\hline
$n\rm L$ & 1S
\\
\hline
$S$ & 1
\\
\hline
$J^{P}$ & $1^{+}$
\\
\hline
$M_{\rm thr.}$ & 9,317
\\
\hline
Our & 9,608
\\
\hline
\cite{QQQQ.ground:all.2018.1}{\scriptsize (VA DA)} & \makecell{$\lesssim$9,390}
\\
\hline
\cite{QQQQ.ground:all.2018.2}{\scriptsize (CM)} & \makecell{9,343, 9,451 {\scriptsize [MM threshold]} \\ 10,174 {\scriptsize [DA, ($\overline 3 \times 3$)-$\rm A\overline S$ mixed with -$\rm A\overline A$ and ($6 \times \overline 6$)-$\rm S\overline A$]} \\ 10,282 {\scriptsize [DA, ($6 \times \overline 6$)-$\rm S\overline A$ mixed with ($\overline 3 \times 3$)-$\rm A\overline S$ and -$\rm A\overline A$]}}
\\
\hline
\cite{QQQQ.ground:all.2019.3}{\scriptsize (NCoQM DA)} & \makecell{9,746 {\scriptsize [($\overline 3 \times 3$)-$\rm S\overline A$]} \\ 9,757 {\scriptsize [($6 \times \overline 6$)-$\rm A\overline S$]}}
\\
\hline
\cite{QQQQ.ground:all.2020.2}{\scriptsize (DMCM DA)} & \makecell{9,610 {\scriptsize [($\overline 3 \times 3$)-$\rm S\overline A$ coupled with -$\rm A\overline A$ and ($6 \times \overline 6$)-$\rm A\overline S$]}}
\\
\hline
\cite{QQQQ.ground:all.exc:ccbb.2021.1}{\scriptsize (DA)} & \makecell{9,343 {\scriptsize [CM, ($\overline 3 \times 3$)-$\rm A\overline S$ coupled with -$\rm A\overline A$ and ($6 \times \overline 6$)-$\rm S\overline A$]} \\ 9,683 {\scriptsize [MCFTM, ($\overline 3 \times 3$)-$\rm A\overline S$ coupled with -$\rm A\overline A$ and ($6 \times \overline 6$)-$\rm S\overline A$]} \\ 9,705 {\scriptsize [MCFTM, ($\overline 3 \times 3$)-$\rm A\overline S$ coupled with -$\rm A\overline A$]} \\ 9,766 {\scriptsize [NCoQM, ($\overline 3 \times 3$)-$\rm A\overline S$ coupled with -$\rm A\overline A$ and ($6 \times \overline 6$)-$\rm S\overline A$]} \\ 9,808 {\scriptsize [NCoQM, ($\overline 3 \times 3$)-$\rm A\overline S$ coupled with -$\rm A\overline A$]}}
\\
\hline
\cite{QQQQ.ground:all.2021.3}{\scriptsize (CM DA)} & \makecell{9,335.1 {\scriptsize [($6 \times \overline 6$)-$\rm S\overline A$ mixed with ($\overline 3 \times 3$)-$\rm A\overline S$ and -$\rm A\overline A$]} \\ 9,498.5 {\scriptsize [($\overline 3 \times 3$)-$\rm A\overline S$ mixed with -$\rm A\overline A$ and ($6 \times \overline 6$)-$\rm S\overline A$]}}
\\
\hline
\cite{QQQQ.ground:all.2021.4}{\scriptsize (NCoQM DA)} & \makecell{9,638 {\scriptsize [($\overline 3 \times 3$)-$\rm A\overline A$ coupled with -$\rm A\overline S$ and ($6 \times \overline 6$)-$\rm S\overline A$]} \\ 9,665 {\scriptsize [($6 \times \overline 6$)-$\rm S\overline A$]} \\ 9,675 {\scriptsize [($\overline 3 \times 3$)-$\rm A\overline S$]}}
\\
\hline
\cite{QQQQ.ground:all.2022.3}{\scriptsize (NCoQM DA)} & \makecell{9,605 {\scriptsize [INS, ($6 \times \overline 6$)-$\rm S\overline A$ mixed with ($\overline 3 \times 3$)-$\rm A\overline S$ and -$\rm A\overline A$]} \\ 9,630 {\scriptsize [INS, ($\overline 3 \times 3$)-$\rm A\overline S$ mixed with -$\rm A\overline A$ and ($6 \times \overline 6$)-$\rm S\overline A$]} \\ 9,676 {\scriptsize [OGE, ($\overline 3 \times 3$)-$\rm A\overline S$ mixed with -$\rm A\overline A$ and ($6 \times \overline 6$)-$\rm S\overline A$]} \\ 9,718 {\scriptsize [OGE, ($6 \times \overline 6$)-$\rm S\overline A$ mixed with ($\overline 3 \times 3$)-$\rm A\overline S$ and -$\rm A\overline A$]}}
\\
\hline
\cite{QQQQ.ground:all.2023.1}{\scriptsize (NCoQM DA)} & \makecell{9,625 {\scriptsize [($6 \times \overline 6$)-$\rm S\overline A$ mixed with ($\overline 3 \times 3$)-$\rm A\overline A$ and -$\rm A\overline S$]} \\ 9,729 {\scriptsize [($\overline 3 \times 3$)-$\rm A\overline S$ mixed with -$\rm A\overline A$ and ($6 \times \overline 6$)-$\rm S\overline A$]}}
\end{tabular}
\end{ruledtabular}
\end{table*}

\begin{table*}
\caption{Same as in Table~\ref{Tab:Comp.cccb.AA.1S}, but for the double charmed and double bottom fully heavy tetraquark ($\rm cc \overline b\overline b$, $\rm bb\overline c\overline c$) with the axialvector--axialvector spin configuration in the ground state. Notations are explained in the caption of Table~\ref{Tab:Comp.cccb.AA.1S}.\label{Tab:Comp.ccbb.AA.1S}}
\begin{ruledtabular}
\begin{tabular}{cccc}
$\rm d\overline d'$ & \multicolumn{3}{c}{$\rm A\overline A$}
\\
\hline
$n\rm L$ & \multicolumn{3}{c}{1S}
\\
\hline
$S$ & 0 & 1 & 2
\\
\hline
$J^{P}$ & $0^{+}$ & $1^{+}$ & $2^{+}$
\\
\hline
$M_{\rm thr.}$ & 12,549 & 12,607 & 12,666
\\
\hline
Our & 12,848 & 12,852 & 12,859
\\
\hline
\cite{QQQQ.ground:all.2018.1}{\scriptsize (VA DA)} & \multicolumn{3}{c}{$\lesssim$12,580}
\\
\hline
\cite{QQQQ.ground:all.2018.2}{\scriptsize (CM)} & \linebreakvertS{12,597 {\scriptsize [MM threshold]} \\ 13,496 {\scriptsize [($\overline 3 \times 3$)-$\rm A\overline A$ mixed with ($6 \times \overline 6$)-$\rm S\overline S$]}} & \linebreakvertS{12,660 {\scriptsize [MM threshold]} \\ 13,560 {\scriptsize [DA, ($\overline 3 \times 3$)]}} & \linebreakvertS{12,695 {\scriptsize [MM threshold]} \\ 13,595 {\scriptsize [DA, ($\overline 3 \times 3$)]}}
\\
\hline
\cite{QQQQ.ground:cbcb;ccbb.2019.1}{\scriptsize (NChQM)} & \linebreakvertS{12,683.9 {\scriptsize [MM, ($1 \times 1$) coupled with ($8 \times 8$)]} \\ 12,683.9 {\scriptsize [mix]} \\ 12,891.5 {\scriptsize [DA, ($\overline 3 \times 3$)-$\rm A\overline A$ coupled with ($6 \times \overline 6$)-$\rm S\overline S$]} \\ 13,140 {\scriptsize [resonance]}} & \linebreakvertS{12,737.4 {\scriptsize [MM, ($1 \times 1$) coupled with ($8 \times 8$)]} \\ 12,737.4 {\scriptsize [mix]} \\ 12,897.6 {\scriptsize [DA, ($\overline 3 \times 3$)]} \\ 13,180 {\scriptsize [resonance]}} & \linebreakvertS{12,790.7 {\scriptsize [MM, ($1 \times 1$) coupled with ($8 \times 8$)]} \\ 12,790.7 {\scriptsize [mix]} \\ 12,904.5 {\scriptsize [DA, ($\overline 3 \times 3$)]} \\ 13,230 {\scriptsize [resonance]}}
\\
\hline
\cite{QQQQ.ground:all.2019.3}{\scriptsize (NCoQM DA)} & 12,953 & 12,960 & 12,972
\\
\hline
\cite{QQQQ.ground;exc:cccc;bbbb;ccbb.2019.4}{\scriptsize (NCoQM DA)} & \linebreakvertS{12,863 {\scriptsize [Model $\RNumb{1}$, ($\overline 3 \times 3$)-$\rm A\overline A$]} \\ 12,866 {\scriptsize [Model $\RNumb{1}$, ($\overline 3 \times 3$)-$\rm A\overline A$ mixed with ($6 \times \overline 6$)-$\rm S\overline S$]} \\ 12,886 {\scriptsize [Model $\RNumb{2}$, ($\overline 3 \times 3$)-$\rm A\overline A$ mixed with ($6 \times \overline 6$)-$\rm S\overline S$]} \\ 12,915 {\scriptsize [Model $\RNumb{2}$, ($\overline 3 \times 3$)-$\rm A\overline A$]}} & \linebreakvertS{12,864 {\scriptsize [Model $\RNumb{1}$]} \\ 12,924 {\scriptsize [Model $\RNumb{2}$]}} & \linebreakvertS{12,868 {\scriptsize [Model $\RNumb{1}$]} \\ 12,940 {\scriptsize [Model $\RNumb{2}$]}}
\\
\hline
\cite{qqQQ.ground:sbsb;ssbb;qqbb.QQQQ.ground;exc:cccc;cbcb;bbbb;ccbb.2020.1}{\scriptsize (RDM DA)} & $12,445 \pm 210$ & 12,536 & 12,614
\\
\hline
\cite{QQQQ.ground:all.2020.2}{\scriptsize (DMCM DA)} & \linebreakvertS{12,865 {\scriptsize [($\overline 3 \times 3$)-$\rm A\overline A$ coupled with ($6 \times \overline 6$)-$\rm S\overline S$]}} & 12,908 & 12,926
\\
\hline
\cite{QQQQ.ground:all.exc:ccbb.2021.1}{\scriptsize (DA)} & \linebreakvertS{12,597 {\scriptsize [CM, ($\overline 3 \times 3$)-$\rm A\overline A$ coupled with ($6 \times \overline 6$)-$\rm S\overline S$]} \\ 12,906 {\scriptsize [MCFTM, ($\overline 3 \times 3$)-$\rm A\overline A$ coupled with ($6 \times \overline 6$)-$\rm S\overline S$]} \\ 12,940 {\scriptsize [MCFTM, ($\overline 3 \times 3$)-$\rm A\overline A$]} \\ 12,963 {\scriptsize [NCoQM, ($\overline 3 \times 3$)-$\rm A\overline A$ coupled with ($6 \times \overline 6$)-$\rm S\overline S$]} \\ 13,023 {\scriptsize [NCoQM, ($\overline 3 \times 3$)-$\rm A\overline A$]}} & \linebreakvertS{12,660 {\scriptsize [CM]} \\ 12,945 {\scriptsize [MCFTM]} \\ 13,024 {\scriptsize [NCoQM]}} & \linebreakvertS{12,695 {\scriptsize [CM]} \\ 12,960 {\scriptsize [MCFTM]} \\ 13,041 {\scriptsize [NCoQM]}}
\\
\hline
\cite{QQQQ.ground:ccbb.2021.2, QQQQ.ground:cccc;cbcb;bbbb;ccbb.2022.1}{\scriptsize (QCDSR DA)} & $12,330^{+180}_{-150}$ & & $12,370^{+190}_{-160}$
\\
\hline
\cite{QQQQ.ground:all.2021.3}{\scriptsize (CM DA)} & \linebreakvertS{12,711.9 {\scriptsize [($\overline 3 \times 3$)-$\rm A\overline A$ mixed with ($6 \times \overline 6$)-$\rm S\overline S$]}} & 12,671.7 & 12,703.1
\\
\hline
\cite{QQQQ.ground:all.2021.4}{\scriptsize (NCoQM)} & \linebreakvertS{12,550 {\scriptsize [MM, ($1 \times 1$) coupled]} \\ 12,853 {\scriptsize [MM, ($8 \times 8$) coupled]} \\ 12,836 {\scriptsize [DA, ($\overline 3 \times 3$)-$\rm A\overline A$ coupled with ($6 \times \overline 6$)-$\rm S\overline S$]} \\ 12,867 {\scriptsize [DA, ($\overline 3 \times 3$)-$\rm A\overline A$]} \\ 12,689, 12,695, 12,832, 12,839 {\scriptsize [other coupled spatial configurations]}} & \linebreakvertS{12,624 {\scriptsize [MM, ($1 \times 1$) coupled]} \\ 12,851 {\scriptsize [MM, ($8 \times 8$) coupled]} \\ 12,878 {\scriptsize [DA, ($\overline 3 \times 3$)]} \\ 12,759, 12,788, 12,878, 12,880 {\scriptsize [other coupled spatial configurations]}} & \linebreakvertS{12,698 {\scriptsize [MM, ($1 \times 1$)]} \\ 12,916 {\scriptsize [MM, ($8 \times 8$)]} \\ 12,899 {\scriptsize [DA, ($\overline 3 \times 3$)-$\rm A\overline A$]} \\ 12,806, 12,828, 12,900, 12,901 {\scriptsize [other coupled spatial configurations]}}
\\
\hline
\cite{QQQQ.ground:cccb;ccbb;bbbc.2022.2}{\scriptsize (DDM DA)} & \linebreakvertS{12,381 {\scriptsize [set $\RNumb{2}$]} \\ 12,401 {\scriptsize [set $\RNumb{1}$]}} & \linebreakvertS{12,390 {\scriptsize [set $\RNumb{2}$]} \\ 12,409 {\scriptsize [set $\RNumb{1}$]}} & \linebreakvertS{12,408 {\scriptsize [set $\RNumb{2}$]} \\ 12,427 {\scriptsize [set $\RNumb{1}$]}}
\\
\hline
\cite{QQQQ.ground:all.2022.3}{\scriptsize (NCoQM DA)} & \linebreakvertS{12,791 {\scriptsize [INS, ($\overline 3 \times 3$)-$\rm A\overline A$ mixed with ($6 \times \overline 6$)-$\rm S\overline S$]} \\ 12,880 {\scriptsize [OGE, ($\overline 3 \times 3$)-$\rm A\overline A$ mixed with ($6 \times \overline 6$)-$\rm S\overline S$]}} & \linebreakvertS{12,818 {\scriptsize [INS]} \\ 12,890 {\scriptsize [OGE]}} & \linebreakvertS{12,838 {\scriptsize [INS]} \\ 12,902 {\scriptsize [OGE]}}
\\
\hline
\cite{QQQQ.ground:all.2023.1}{\scriptsize (NCoQM DA)} & \linebreakvertS{12,920 {\scriptsize [($\overline 3 \times 3$)-$\rm A\overline A$ mixed with ($6 \times \overline 6$)-$\rm S\overline S$]}} & 12,940 & 12,961
\end{tabular}
\end{ruledtabular}
\end{table*}

\begin{table*}
\caption{Same as in Table~\ref{Tab:Comp.ccbb.AA.1S}, but for the first orbital excitation. Notations are explained in the caption of Table~\ref{Tab:Comp.cccb.AA.1S}.\label{Tab:Comp.ccbb.AA.1P}}
\begin{ruledtabular}
\begin{tabular}{cccccccc}
$\rm d\overline d'$ & \multicolumn{7}{c}{$\rm A\overline A$}
\\
\hline
$n\rm L$ & \multicolumn{7}{c}{1P}
\\
\hline
$S$ & 1 & 0 & 1 & 2 & 1 & 2 & 2
\\
\hline
$J^{P}$ & $0^{-}$ & \multicolumn{3}{c}{$1^{-}$} & \multicolumn{2}{c}{$2^{-}$} & $3^{-}$
\\
\hline
$M_{\rm thr.}$ & 12,973 & \multicolumn{3}{c}{13,017} & \multicolumn{2}{c}{13,035} & 13,094
\\
\hline
Our & 13,106 & 13,103 & 13,108 & 13,111 & 13,106 & 13,112 & 13,110
\\
\hline 
\cite{qqQQ.ground:sbsb;ssbb;qqbb.QQQQ.ground;exc:cccc;cbcb;bbbb;ccbb.2020.1}{\scriptsize (RDM DA)} & 12,976 & 12,967 & & 12,977 & & &
\\
\hline
\cite{QQQQ.ground:all.exc:ccbb.2021.1}{\scriptsize (DA)} & & \linebreakvertP{13,204 {\scriptsize [MCFTM, ($\overline 3 \times 3$)-$\rm A\overline A$]} \\ 13,204 {\scriptsize [MCFTM, ($\overline 3 \times 3$)-$\rm A\overline A$ coupled with ($6 \times \overline 6$)-$\rm S\overline S$]}} & & & & &
\end{tabular}
\end{ruledtabular}
\end{table*}

\begin{table*}
\caption{Same as in Table~\ref{Tab:Comp.ccbb.AA.1S}, but for the first radial excitation. Notations are explained in the caption of Table~\ref{Tab:Comp.cccb.AA.1S}.\label{Tab:Comp.ccbb.AA.2S}}
\begin{ruledtabular}
\begin{tabular}{cccc}
$\rm d\overline d'$ & \multicolumn{3}{c}{$\rm A\overline A$}
\\
\hline
$n\rm L$ & \multicolumn{3}{c}{2S}
\\
\hline
$S$ & 0 & 1 & 2
\\
\hline
$J^{P}$ & $0^{+}$ & $1^{+}$ & $2^{+}$
\\
\hline
$M_{\rm thr.}$ & 12,549 & 12,607 & 12,666
\\
\hline
Our & 13,282 & 13,282 & 13,283
\\
\hline
\cite{QQQQ.ground;exc:cccc;bbbb;ccbb.2019.4}{\scriptsize (NCoQM DA)} & & \linebreakvertS{13,259 {\scriptsize [Model $\RNumb{1}$]} \\ 13,321 {\scriptsize [Model $\RNumb{2}$]}} & \linebreakvertS{13,262 {\scriptsize [Model $\RNumb{1}$]} \\ 13,334 {\scriptsize [Model $\RNumb{2}$]}}
\\
\hline
\cite{qqQQ.ground:sbsb;ssbb;qqbb.QQQQ.ground;exc:cccc;cbcb;bbbb;ccbb.2020.1}{\scriptsize (RDM DA)} & 13,017 & 13,060 & 13,101
\end{tabular}
\end{ruledtabular}
\end{table*}

\begin{table*}
\caption{Same as in Table~\ref{Tab:Comp.ccbb.AA.1S}, but for the second orbital excitation. Notations are explained in the caption of Table~\ref{Tab:Comp.cccb.AA.1S}.\label{Tab:Comp.ccbb.AA.1D}}
\begin{ruledtabular}
\begin{tabular}{cccccccccc}
$\rm d\overline d'$ & \multicolumn{9}{c}{$\rm A\overline A$}
\\
\hline
$n\rm L$ & \multicolumn{9}{c}{1D}
\\
\hline
$S$ & 2 & 1 & 2 & 0 & 1 & 2 & 1 & 2 & 2
\\
\hline
$J^{P}$ & $0^{+}$ & \multicolumn{2}{c}{$1^{+}$} & \multicolumn{3}{c}{$2^{+}$} & \multicolumn{2}{c}{$3^{+}$} & $4^{+}$
\\
\hline
$M_{\rm thr.}$ & 12,549 & \multicolumn{2}{c}{12,607} & \multicolumn{3}{c}{12,666} & \multicolumn{2}{c}{13,303} & 13,362
\\
\hline
Our & 13,330 & 13,328 & 13,331 & 13,324 & 13,330 & 13,334 & 13,327 & 13,332 & 13,329
\\
\hline 
\cite{qqQQ.ground:sbsb;ssbb;qqbb.QQQQ.ground;exc:cccc;cbcb;bbbb;ccbb.2020.1}{\scriptsize (RDM DA)} & 13,208 & 13,205 & 13,206 & 13,204 & & 13,204 & & &
\\
\hline
\cite{QQQQ.ground:all.exc:ccbb.2021.1}{\scriptsize (DA)} & & & & \linebreakvertD{13,398 {\scriptsize [MCFTM, ($\overline 3 \times 3$)-$\rm A\overline A$]} \\ 13,398 {\scriptsize [MCFTM, ($\overline 3 \times 3$)-$\rm A\overline A$ coupled with ($6 \times \overline 6$)-$\rm S\overline S$]}} & & & & &
\end{tabular}
\end{ruledtabular}
\end{table*}

\begin{table*}
\caption{Same as in Table~\ref{Tab:Comp.ccbb.AA.1S}, but for the first orbital-radial excitation. Notations are explained in the caption of Table~\ref{Tab:Comp.cccb.AA.1S}.\label{Tab:Comp.ccbb.AA.2P}}
\begin{ruledtabular}
\begin{tabular}{cccccccc}
$\rm d\overline d'$ & \multicolumn{7}{c}{$\rm A\overline A$}
\\
\hline
$n\rm L$ & \multicolumn{7}{c}{2P}
\\
\hline
$S$ & 1 & 0 & 1 & 2 & 1 & 2 & 2
\\
\hline
$J^{P}$ & $0^{-}$ & \multicolumn{3}{c}{$1^{-}$} & \multicolumn{2}{c}{$2^{-}$} & $3^{-}$
\\
\hline
$M_{\rm thr.}$ & 12,973 & \multicolumn{3}{c}{13,017} & \multicolumn{2}{c}{13,035} & 13,094
\\
\hline
Our & 13,468 & 13,461 & 13,468 & 13,472 & 13,463 & 13,470 & 13,466
\\
\hline
\cite{qqQQ.ground:sbsb;ssbb;qqbb.QQQQ.ground;exc:cccc;cbcb;bbbb;ccbb.2020.1}{\scriptsize (RDM DA)} & 13,311 & 13,304 & & 13,311 & & &
\end{tabular}
\end{ruledtabular}
\end{table*}

\begin{table*}
\caption{Same as in Table~\ref{Tab:Comp.ccbb.AA.1S}, but for the second radial excitation. Notations are explained in the caption of Table~\ref{Tab:Comp.cccb.AA.1S}.\label{Tab:Comp.ccbb.AA.3S}}
\begin{ruledtabular}
\begin{tabular}{cccc}
$\rm d\overline d'$ & \multicolumn{3}{c}{$\rm A\overline A$}
\\
\hline
$n\rm L$ & \multicolumn{3}{c}{3S}
\\
\hline
$S$ & 0 & 1 & 2
\\
\hline
$J^{P}$ & $0^{+}$ & $1^{+}$ & $2^{+}$
\\
\hline
$M_{\rm thr.}$ & 12,549 & 12,607 & 12,666
\\
\hline
Our & 13,629 & 13,629 & 13,628
\\
\hline 
\cite{QQQQ.ground;exc:cccc;bbbb;ccbb.2019.4}{\scriptsize (NCoQM DA)} & & \linebreakvertS{13,297 {\scriptsize [Model $\RNumb{1}$]} \\ 13,364 {\scriptsize [Model $\RNumb{2}$]}} & \linebreakvertS{13,299 {\scriptsize [Model $\RNumb{1}$]} \\ 13,375 {\scriptsize [Model $\RNumb{2}$]}}
\\
\hline
\cite{qqQQ.ground:sbsb;ssbb;qqbb.QQQQ.ground;exc:cccc;cbcb;bbbb;ccbb.2020.1}{\scriptsize (RDM DA)} & 13,349 & 13,381 & 13,412
\end{tabular}
\end{ruledtabular}
\end{table*}

\begin{table*}
\caption{Same as in Table~\ref{Tab:Comp.cccb.AA.1S}, but for the triple bottom and charmed fully heavy tetraquark ($\rm bb\overline b\overline c$, $\rm cb\overline b\overline b$) with the axialvector--axialvector spin configuration in the ground state. Notations are explained in the caption of Table~\ref{Tab:Comp.cccb.AA.1S}.\label{Tab:Comp.bbbc.AA.1S}}
\begin{ruledtabular}
\begin{tabular}{cccc}
$\rm d\overline d'$ & \multicolumn{3}{c}{$\rm A\overline A$}
\\
\hline
$n\rm L$ & \multicolumn{3}{c}{1S}
\\
\hline
$S$ & 0 & 1 & 2
\\
\hline
$J^{P}$ & $0^{+}$ & $1^{+}$ & $2^{+}$
\\
\hline
$M_{\rm thr.}$ & 15,673 & 15,732 & 15,793
\\
\hline
Our & 16,102 & 16,104 & 16,108
\\
\hline
\cite{QQQQ.ground:all.2018.1}{\scriptsize (VA DA)} & \multicolumn{3}{c}{$\lesssim$15,770}
\\
\hline
\cite{QQQQ.ground:all.2018.2}{\scriptsize (CM)} & \linebreakvertS{15,713 {\scriptsize [MM threshold]} \\ 16,823 {\scriptsize [DA, ($\overline 3 \times 3$)-$\rm A\overline A$ mixed with ($6 \times \overline 6$)-$\rm S\overline S$]}} & \linebreakvertS{15,773 {\scriptsize [MM threshold]} \\ 16,884 {\scriptsize [DA, ($\overline 3 \times 3$)-$\rm A\overline A$ mixed with -$\rm A\overline S$ and ($6 \times \overline 6$)-$\rm S\overline A$]}} &  \linebreakvertS{15,806 {\scriptsize [MM threshold]} \\ 16,917 {\scriptsize [DA, ($\overline 3 \times 3$)]}}
\\
\hline
\cite{QQQQ.ground:all.2019.3}{\scriptsize (NCoQM DA)} & 16,158 & 16,164 & 16,176
\\
\hline
\cite{QQQQ.ground:all.2020.2}{\scriptsize (DMCM DA)} & \linebreakvertS{16,040 {\scriptsize [($\overline 3 \times 3$)-$\rm A\overline A$ coupled with ($6 \times \overline 6$)-$\rm S\overline S$]}} & \linebreakvertS{16,013 {\scriptsize [($\overline 3 \times 3$)-$\rm A\overline A$ coupled with -$\rm S\overline A$ and ($6 \times \overline 6$)-$\rm A\overline S$]}} & 16,129
\\
\hline
\cite{QQQQ.ground:all.exc:ccbb.2021.1}{\scriptsize (DA)} & \linebreakvertS{15,713 {\scriptsize [CM, ($\overline 3 \times 3$)-$\rm A\overline A$ coupled with ($6 \times \overline 6$)-$\rm S\overline S$]} \\ 16,126 {\scriptsize [MCFTM, ($\overline 3 \times 3$)-$\rm A\overline A$ coupled with ($6 \times \overline 6$)-$\rm S\overline S$]} \\ 16,158 {\scriptsize [MCFTM, ($\overline 3 \times 3$)-$\rm A\overline A$]} \\ 16,175 {\scriptsize [NCoQM, ($\overline 3 \times 3$)-$\rm A\overline A$ coupled with ($6 \times \overline 6$)-$\rm S\overline S$]} \\ 16,224 {\scriptsize [NCoQM, ($\overline 3 \times 3$)-$\rm A\overline A$]}} & \linebreakvertS{15,729 {\scriptsize [CM, ($\overline 3 \times 3$)-$\rm A\overline A$ coupled with -$\rm A\overline S$ and ($6 \times \overline 6$)-$\rm S\overline A$]} \\ 16,130 {\scriptsize [MCFTM, ($\overline 3 \times 3$)-$\rm A\overline A$ coupled with -$\rm A\overline S$ and ($6 \times \overline 6$)-$\rm S\overline A$]} \\ 16,151 {\scriptsize [MCFTM, ($\overline 3 \times 3$)-$\rm A\overline A$ coupled with -$\rm A\overline S$]} \\ 16,179 {\scriptsize [NCoQM, ($\overline 3 \times 3$)-$\rm A\overline A$ coupled with -$\rm A\overline S$ and ($6 \times \overline 6$)-$\rm S\overline A$]} \\ 16,230 {\scriptsize [NCoQM, ($\overline 3 \times 3$)-$\rm A\overline A$ coupled with -$\rm A\overline S$]}} & \linebreakvertS{15,806 {\scriptsize [CM]} \\ 16,182 {\scriptsize [MCFTM]} \\ 16,274 {\scriptsize [NCoQM]}}
\\
\hline
\cite{QQQQ.ground:all.2021.3}{\scriptsize (CM DA)} & \linebreakvertS{15,862.0 {\scriptsize [($\overline 3 \times 3$)-$\rm A\overline A$ mixed with ($6 \times \overline 6$)-$\rm S\overline S$]}} & \linebreakvertS{15,851.3 {\scriptsize [($\overline 3 \times 3$)-$\rm A\overline A$ mixed with -$\rm A\overline S$ and ($6 \times \overline 6$)-$\rm S\overline A$]}} & 15,882.3
\\
\hline
\cite{QQQQ.ground:all.2021.4}{\scriptsize (NCoQM)} & \linebreakvertS{15,676 {\scriptsize [MM, ($1 \times 1$) coupled]} \\ 15,976 {\scriptsize [MM, ($8 \times 8$) coupled]} \\ 16,012 {\scriptsize [DA, ($\overline 3 \times 3$)-$\rm A\overline A$ coupled with ($6 \times \overline 6$)-$\rm S\overline S$]} \\ 16,058 {\scriptsize [DA, ($\overline 3 \times 3$)-$\rm A\overline A$]} \\ 15,892, 15,906, 15,977, 16,013 {\scriptsize [other coupled spatial configurations]}} & \linebreakvertS{15,738 {\scriptsize [MM, ($1 \times 1$) coupled]} \\ 15,993 {\scriptsize [MM, ($8 \times 8$) coupled]} \\ 16,017 {\scriptsize [DA, ($\overline 3 \times 3$)-$\rm A\overline A$ coupled with -$\rm A\overline S$ and ($6 \times \overline 6$)-$\rm S\overline A$]} \\ 16,068 {\scriptsize [DA, ($\overline 3 \times 3$)-$\rm A\overline A$]} \\ 15,922, 15,944, 15,985, 16,018 {\scriptsize [other coupled coupled spatial configurations}} & \linebreakvertS{15,812 {\scriptsize [MM, ($1 \times 1$)]} \\ 16,030 {\scriptsize [MM, ($8 \times 8$)]} \\ 16,087 {\scriptsize [DA, ($\overline 3 \times 3$)]} \\ 15,980, 16,008, 16,065, 16,087 {\scriptsize [other coupled spatial configurations]}}
\\
\hline
\cite{QQQQ.ground:cccb;ccbb;bbbc.2022.2}{\scriptsize (DDM DA)} & \linebreakvertS{16,049 {\scriptsize [set $\RNumb{2}$]} \\ 16,060 {\scriptsize [set $\RNumb{1}$]}} & \linebreakvertS{16,052 {\scriptsize [set $\RNumb{2}$]} \\ 16,062 {\scriptsize [set $\RNumb{1}$]}} & \linebreakvertS{16,058 {\scriptsize [set $\RNumb{2}$]} \\ 16,068 {\scriptsize [set $\RNumb{1}$]}}
\\
\hline
\cite{QQQQ.ground:all.2022.3}{\scriptsize (NCoQM DA)} & \linebreakvertS{16,019 {\scriptsize [INS, ($\overline 3 \times 3$)-$\rm A\overline A$ mixed with ($6 \times \overline 6$)-$\rm S\overline S$]} \\ 16,061 {\scriptsize [OGE, ($\overline 3 \times 3$)-$\rm A\overline A$ mixed with ($6 \times \overline 6$)-$\rm S\overline S$]}} & \linebreakvertS{16,053 {\scriptsize [INS, ($\overline 3 \times 3$)-$\rm A\overline A$ mixed with -$\rm A\overline S$ and ($6 \times \overline 6$)-$\rm S\overline A$]} \\ 16,079 {\scriptsize [OGE, ($\overline 3 \times 3$)-$\rm A\overline A$ mixed with -$\rm A\overline S$ and ($6 \times \overline 6$)-$\rm S\overline A$]}} & \linebreakvertS{16,051 {\scriptsize [INS]} \\ 16,089 {\scriptsize [OGE]}}
\\
\hline
\cite{QQQQ.ground:all.2023.1}{\scriptsize (NCoQM DA)} & \linebreakvertS{16,163 {\scriptsize [($\overline 3 \times 3$)-$\rm A\overline A$ mixed with ($6 \times \overline 6$)-$\rm S\overline S$]}} & \linebreakvertS{16,125 {\scriptsize [($\overline 3 \times 3$)-$\rm A\overline A$ mixed with -$\rm A\overline S$ and ($6 \times \overline 6$)-$\rm S\overline A$]}} & 16,149
\end{tabular}
\end{ruledtabular}
\end{table*}

\begin{table*}
\caption{Same as in Table~\ref{Tab:Comp.bbbc.AA.1S}, but for the scalar--axialvector spin configuration in ground state. Notations are explained in the caption of Table~\ref{Tab:Comp.cccb.AA.1S}.\label{Tab:Comp.bbbc.SA.1S}}
\begin{ruledtabular}
\begin{tabular}{cc}
$\rm d\overline d'$ & $\rm S\overline A$, $\rm A\overline S$
\\
\hline
$n\rm L$ & 1S
\\
\hline
$S$ & 1
\\
\hline
$J^{P}$ & $1^{+}$
\\
\hline
$M_{\rm thr.}$ & 15,732
\\
\hline
Our & 16,099
\\
\hline
\cite{QQQQ.ground:all.2018.1}{\scriptsize (VA DA)} & \makecell{$\lesssim$15,770}
\\
\hline
\cite{QQQQ.ground:all.2018.2}{\scriptsize (CM)} & \makecell{15,729, 15,804 {\scriptsize [MM threshold]} \\ 16,840 {\scriptsize [DA, ($\overline 3 \times 3$)-$\rm A\overline S$ mixed with -$\rm A\overline A$ and ($6 \times \overline 6$)-$\rm S\overline A$]} \\ 16,915 {\scriptsize [DA, ($6 \times \overline 6$)-$\rm S\overline A$ mixed with ($\overline 3 \times 3$)-$\rm A\overline S$ and -$\rm A\overline A$]}}
\\
\hline
\cite{QQQQ.ground:all.2019.3}{\scriptsize (NCoQM DA)} & \makecell{16,157 {\scriptsize [($\overline 3 \times 3$)-$\rm S\overline A$]} \\ 16,167 {\scriptsize [($6 \times \overline 6$)-$\rm A\overline S$]}}
\\
\hline
\cite{QQQQ.ground:all.2020.2}{\scriptsize (DMCM DA)} & \makecell{16,013 {\scriptsize [($\overline 3 \times 3$)-$\rm S\overline A$ coupled with -$\rm A\overline A$ and ($6 \times \overline 6$)-$\rm A\overline S$]}}
\\
\hline
\cite{QQQQ.ground:all.exc:ccbb.2021.1}{\scriptsize (DA)} & \makecell{15,729 {\scriptsize [CM, ($\overline 3 \times 3$)-$\rm A\overline S$ coupled with -$\rm A\overline A$ and ($6 \times \overline 6$)-$\rm S\overline A$]} \\ 16,130 {\scriptsize [MCFTM, ($\overline 3 \times 3$)-$\rm A\overline S$ coupled with -$\rm A\overline A$ and ($6 \times \overline 6$)-$\rm S\overline A$]} \\ 16,151 {\scriptsize [MCFTM, ($\overline 3 \times 3$)-$\rm A\overline S$ coupled with -$\rm A\overline A$]} \\ 16,179 {\scriptsize [NCoQM, ($\overline 3 \times 3$)-$\rm A\overline S$ coupled with -$\rm A\overline A$ and ($6 \times \overline 6$)-$\rm S\overline A$]} \\ 16,230 {\scriptsize [NCoQM, ($\overline 3 \times 3$)-$\rm A\overline S$ coupled with -$\rm A\overline A$]}}
\\
\hline
\cite{QQQQ.ground:all.2021.3}{\scriptsize (CM DA)} & \makecell{15,719.1 {\scriptsize [($6 \times \overline 6$)-$\rm S\overline A$ mixed with ($\overline 3 \times 3$)-$\rm A\overline S$ and -$\rm A\overline A$]} \\ 15,854.4 {\scriptsize [($\overline 3 \times 3$)-$\rm A\overline S$ mixed with -$\rm A\overline A$ and ($6 \times \overline 6$)-$\rm S\overline A$]}}
\\
\hline
\cite{QQQQ.ground:all.2021.4}{\scriptsize (NCoQM DA)} & \makecell{16,017 {\scriptsize [($\overline 3 \times 3$)-$\rm A\overline A$ coupled with -$\rm A\overline S$ and ($6 \times \overline 6$)-$\rm S\overline A$]} \\ 16,032 {\scriptsize [($6 \times \overline 6$)-$\rm S\overline A$]} \\ 16,049 {\scriptsize [($\overline 3 \times 3$)-$\rm A\overline S$]}}
\\
\hline
\cite{QQQQ.ground:all.2022.3}{\scriptsize (NCoQM DA)} & \makecell{16,009 {\scriptsize [INS, ($6 \times \overline 6$)-$\rm S\overline A$ mixed with ($\overline 3 \times 3$)-$\rm A\overline S$ and -$\rm A\overline A$]} \\ 16,036 {\scriptsize [INS, ($\overline 3 \times 3$)-$\rm A\overline S$ mixed with -$\rm A\overline A$ and ($6 \times \overline 6$)-$\rm S\overline A$]} \\ 16,046 {\scriptsize [OGE, ($\overline 3 \times 3$)-$\rm A\overline S$ mixed with -$\rm A\overline A$ and ($6 \times \overline 6$)-$\rm S\overline A$]} \\ 16,089 {\scriptsize [OGE, ($6 \times \overline 6$)-$\rm S\overline A$ mixed with ($\overline 3 \times 3$)-$\rm A\overline S$ and -$\rm A\overline A$]}}
\\
\hline
\cite{QQQQ.ground:all.2023.1}{\scriptsize (NCoQM DA)} & \makecell{16,043 {\scriptsize [($6 \times \overline 6$)-$\rm S\overline A$ mixed with ($\overline 3 \times 3$)-$\rm A\overline A$ and -$\rm A\overline S$]} \\ 16,144 {\scriptsize [($\overline 3 \times 3$)-$\rm A\overline S$ mixed with -$\rm A\overline A$ and ($6 \times \overline 6$)-$\rm S\overline A$]}}
\end{tabular}
\end{ruledtabular}
\end{table*}

\par
For the further analysis we additionally distinguish results obtained in the following models
\begin{itemize}

	\item diquark--antidiquark picture \cite{QQQQ.ground:all.2018.1, QQQQ.ground:all.2018.2, QQQQ.ground:cbcb;ccbb.2019.1, QQQQ.ground:all.2019.3, QQQQ.ground;exc:cccc;bbbb;ccbb.2019.4, qqQQ.ground:sbsb;ssbb;qqbb.QQQQ.ground;exc:cccc;cbcb;bbbb;ccbb.2020.1, QQQQ.ground:all.2020.2, QQQQ.ground:all.exc:ccbb.2021.1, QQQQ.ground:ccbb.2021.2, QQQQ.ground:all.2021.3, QQQQ.ground:all.2021.4, QQQQ.ground:cccc;cbcb;bbbb;ccbb.2022.1, QQQQ.ground:cccb;ccbb;bbbc.2022.2, QQQQ.ground:all.2022.3, QQQQ.ground:all.2023.1};
	
	\item meson--meson model \cite{QQQQ.ground:cbcb;ccbb.2019.1, QQQQ.ground:all.2021.4};
	
	\item mixture of those two \cite{QQQQ.ground:cbcb;ccbb.2019.1}.
	
\end{itemize}
\par
We now can analyze how our predictions relate to those of the other scientific groups. From the Tables~\ref{Tab:Comp.cccb.AA.1S}-\ref{Tab:Comp.bbbc.SA.1S} it can be seen that our results agree within the $\pm 75$ MeV range with the following predictions. 
\begin{itemize}

	\item For the $\rm cc\overline c\overline b$, $\rm bc\overline c\overline c$ tetraquarks.
	\begin{itemize}
	
		\item In the diquark–-antidiquark pure ($\overline 3 \times 3$)-color configuration: 
		all predictions \cite{QQQQ.ground:cccb;ccbb;bbbc.2022.2};
		only $J^{P}=0,1^{+}$ \cite{QQQQ.ground:all.2021.4};
		$J^{P}=2^{+}$ \cite{QQQQ.ground:all.2022.3}~[INS].
		
		\item In the diquark–-antidiquark ($\overline 3 \times 3$)- and ($6 \times \overline 6$)-color configurations mixed or coupled for $J^{P}=0,1^{+}$:
		all predictions \cite{QQQQ.ground:all.2020.2, QQQQ.ground:all.2021.4}, \cite{QQQQ.ground:all.exc:ccbb.2021.1}~[MCFTM], \cite{QQQQ.ground:all.2022.3}~[INS];
		only $J^{P}=0^{+}$ $A\overline A$- and $J^{P}=1^{+}$ $\rm S\overline A$-, $\rm A\overline S$-spin configurations \cite{QQQQ.ground:all.2022.3}~[OGE];
		only $\rm S\overline A$-, $\rm A\overline S$-spin configurations \cite{QQQQ.ground:all.2023.1}.
		
		\item In the other models: 
		all predictions \cite{QQQQ.ground:all.2021.4} [meson--meson ($8 \times 8$)-color configuration coupled, some other spatial configurations]; 
		only $\rm S\overline A$-, $\rm A\overline S$-spin configurations \cite{QQQQ.ground:all.2021.4} [diquark–-antidiquark pure ($6 \times \overline 6$)-color configuration].
			
	\end{itemize}
	
	\item For the $\rm cc\overline b\overline b$, $\rm bb\overline c\overline c$ tetraquarks.
	\begin{itemize}
	
		\item In the diquark–-antidiquark pure ($\overline 3 \times 3$)-color configuration: 
		1S, all predictions \cite{QQQQ.ground:cbcb;ccbb.2019.1, QQQQ.ground:all.2020.2, QQQQ.ground:all.2021.4, QQQQ.ground:all.2022.3};
		1S, 2S, all predictions \cite{QQQQ.ground;exc:cccc;bbbb;ccbb.2019.4}.
		
		\item In the diquark–-antidiquark ($\overline 3 \times 3$)- and ($6 \times \overline 6$)-color configurations mixed or coupled for $J^{P}=0^{+}$:
		1S, all predictions \cite{QQQQ.ground:cbcb;ccbb.2019.1, QQQQ.ground;exc:cccc;bbbb;ccbb.2019.4, QQQQ.ground:all.2020.2, QQQQ.ground:all.2021.4, QQQQ.ground:all.2022.3, QQQQ.ground:all.2023.1}, \cite{QQQQ.ground:all.exc:ccbb.2021.1}~[MCFTM].
		
		\item In the other models: 
		all predictions \cite{QQQQ.ground:all.2021.4} [meson--meson ($8 \times 8$)-color configuration coupled, some other spatial configurations].	
		
	\end{itemize}
	
	\item For the $\rm bb\overline b\overline c$, $\rm cb\overline b\overline b$ tetraquarks.
	\begin{itemize}
	
		\item In the diquark–-antidiquark pure ($\overline 3 \times 3$)-color configuration: 
		all predictions \cite{QQQQ.ground:all.2019.3, QQQQ.ground:all.2021.4, QQQQ.ground:cccb;ccbb;bbbc.2022.2}, \cite{QQQQ.ground:all.exc:ccbb.2021.1}~[MCFTM];
		$J^{P}=2^{+}$ \cite{QQQQ.ground:all.2020.2, QQQQ.ground:all.2022.3, QQQQ.ground:all.2023.1}.
		
		\item In the diquark–-antidiquark ($\overline 3 \times 3$)- and ($6 \times \overline 6$)-color configurations mixed or coupled for $J^{P}=0,1^{+}$:
		all predictions \cite{QQQQ.ground:all.2023.1}, \cite{QQQQ.ground:all.exc:ccbb.2021.1}~[MCFTM, NCoQM], \cite{QQQQ.ground:all.2022.3}~[OGE];
		only $J^{P}=0^{+}$ \cite{QQQQ.ground:all.2020.2};
		only $J^{P}=1^{+}$ \cite{QQQQ.ground:all.2022.3}~[INS].
		
		\item In the other models: 
		only $J^{P}=2^{+}$ \cite{QQQQ.ground:all.2021.4} [some other spatial configurations]; 
		only $\rm S\overline A$-, $\rm A\overline S$-spin configurations \cite{QQQQ.ground:all.2019.3, QQQQ.ground:all.2021.4} [diquark–-antidiquark pure ($6 \times \overline 6$)-color configuration].
			
	\end{itemize}
	
\end{itemize}
\par
A number of other conclusions can be drawn from this comparison.
\begin{itemize}

	\item Predictions of Ref.~\cite{QQQQ.ground;exc:cccc;bbbb;ccbb.2019.4} (in particular, Model $\RNumb{1}$) are in a good agreement with our results for the ground state $\rm cc\overline b\overline b$, $\rm bb\overline c\overline c$ tetraquarks in the diquark--antidiquark picture;
	
	\item Predictions of Refs.~\cite{QQQQ.ground:all.2020.2, QQQQ.ground:all.2021.4} give partial agreement for the ground state $\rm cc\overline c\overline b$, $\rm bc\overline c\overline c$ and $\rm cc\overline b\overline b$, $\rm bb\overline c\overline c$ tetraquarks in the diquark--antidiquark picture;
	
	\item Predictions of Refs.~\cite{QQQQ.ground:all.2022.3} (in particular, INS) give partial agreement for the ground state $\rm cc\overline c\overline b$, $\rm bc\overline c\overline c$ tetraquarks in the diquark--antidiquark picture;
	
	\item Predictions of Ref.~\cite{QQQQ.ground:all.2021.4} give partial agreement for the ground state $\rm cc\overline b\overline b$, $\rm bb\overline c\overline c$ tetraquarks in the meson--meson model;
	
	\item There are no predictions which are in good agreement for the ground state $\rm bb\overline b\overline c$, $\rm cb\overline b\overline b$ tetraquarks.
	
\end{itemize}
\par
In addition, the comparison of our results with those of other scientific groups shows the following.
\begin{itemize}

	\item For the $\rm cc\overline c\overline b$, $\rm bc\overline c\overline c$ tetraquarks, our results in general are slightly lighter than those of most other scientific groups;
	
	\item For the $\rm cc\overline b\overline b$, $\rm bb\overline c\overline c$ tetraquarks, our results are generally median (there are many results giving both heavier and lighter masses);
	
	\item For the $\rm bb\overline b\overline c$, $\rm cb\overline b\overline b$ tetraquarks, our results in general are slightly heavier than those of most other scientific groups.
	
\end{itemize}
\par
It is interesting, that the conclusions of different groups are sometimes quite contradictory. Namely, the following outlooks are made.
\begin{itemize}

	\item For the $\rm cc\overline c\overline b$, $\rm bc\overline c\overline c$ tetraquarks.
	\begin{itemize}
	
		\item In Ref.~\cite{qqqq;qqqQ;qqQQ;qQQQ;QQQQ.ground:all.1992.1} there is an indication of possible weak binding for the ground $J^{P}=1^{+}$ state (see Ref.~\cite{qqqq;qqqQ;qqQQ;qQQQ;QQQQ.ground:all.1992.1}, Fig.~8). The authors of Ref.~\cite{QQQQ.ground:all.2021.4} predict a number of narrow resonances with fall-apart strong decay widths lying in the range of $1-10$ MeV (see Ref.~\cite{QQQQ.ground:all.2021.4}, Table~$\RNumb{12}$). Additionally, in Refs.~\cite{QQQQ.ground:all.2020.2, QQQQ.ground:all.exc:ccbb.2021.1, QQQQ.ground:cccb;ccbb;bbbc.2022.2} the tetraquark is predicted to be compact.
		
		\item While the authors of Refs.~\cite{QQQQ.ground:all.2019.3, QQQQ.ground;exc:cccc;bbbb;ccbb.2019.4, QQQQ.ground:all.2020.2, QQQQ.ground:all.2021.3, QQQQ.ground:cccb;ccbb;bbbc.2022.2, QQQQ.ground:all.2023.1} come to the conclusion that narrow bound ground states are not favored since all of the masses lie significantly above the lowest fall-apart strong decay thresholds. Yet there is a possibility for the resonances. Note, that in Ref.~\cite{QQQQ.ground:all.2018.2} it is stated that ``if this state does exist, it should be less stable than the lowest $J^{P}=1^{+}$ $\rm bb\overline b\overline c$''. Also in Ref.~\cite{QQQQ.ground:all.2020.2} the $J^{P}=0,1^{+}$ ground states are almost degenerate and the $J^{P}=2^{+}$ ground state lies $\approx$100 MeV above them.
		
	\end{itemize}
	
	\item For the $\rm cc\overline b\overline b$, $\rm bb\overline c\overline c$ tetraquarks.
	\begin{itemize}

		\item The authors of Refs.~\cite{qqQQ.ground:sbsb;ssbb;qqbb.QQQQ.ground;exc:cccc;cbcb;bbbb;ccbb.2020.1, QQQQ.ground:ccbb.2021.2, QQQQ.ground:cccc;cbcb;bbbb;ccbb.2022.1, QQQQ.ground:cccb;ccbb;bbbc.2022.2} conclude, that the ground (S-wave positive parity) states lie below the fall-apart strong decay thresholds, implying that these tetraquark states can only undergo radiative transitions or weak decays and, thus, are expected to be narrow. Note that in Ref.~\cite{QQQQ.ground:all.2018.1} the authors argue that this tetraquark ``remains stable against strong decay'', because it cannot ``decay strongly by annihilating at least a pair of quarks and antiquarks of the same flavor''. Additionally, the authors of Ref.~\cite{QQQQ.ground:all.2021.4} predict a number of narrow resonances with fall-apart strong decay widths lying in the range of $1-20$ MeV (see Ref.~\cite{QQQQ.ground:all.2021.4}, Table~$\RNumb{20}$). Also, in Refs.~\cite{QQQQ.ground:all.2020.2, QQQQ.ground:all.exc:ccbb.2021.1, QQQQ.ground:cccb;ccbb;bbbc.2022.2, QQQQ.ground:all.2023.1} the tetraquark is predicted to be compact.
		
		\item At the same time the authors of Refs.~\cite{qqqq;qqqQ;qqQQ;qQQQ;QQQQ.ground:all.1992.1, qqQQ:udbb;ssbb.QQQQ:ccbb.2015.1, tetrons2018, QQQQ.ground:all.2018.2, QQQQ.ground:cbcb;ccbb.2019.1, QQQQ.ground:all.2019.3, QQQQ.ground;exc:cccc;bbbb;ccbb.2019.4, QQQQ.ground:all.2020.2, QQQQ.ground:all.2021.3, QQQQ.ground:all.2023.1} argue that these tetraquarks should not be narrow and stable or should not be bound and exist at all~\cite{qqqq;qqqQ;qqQQ;qQQQ;QQQQ.ground:all.1992.1, qqQQ:udbb;ssbb.QQQQ:ccbb.2015.1}. In Refs.~\cite{QQQQ.ground:cbcb;ccbb.2019.1, QQQQ.ground:all.2019.3, QQQQ.ground;exc:cccc;bbbb;ccbb.2019.4, QQQQ.ground:all.2020.2, QQQQ.ground:all.2021.3} the masses in the ground states are all much lager than the relevant fall-apart strong decay thresholds indicating possibility only for the resonances. Moreover, the authors of Ref.~\cite{QQQQ:ccbb.2019.2} find a rather short lifetime $\tau(\rm X_{cc\overline b\overline b}) = (0.1-0.3) \times 10^{-12}$s which is approximately 3 times smaller that that of the $\rm B_{c}$-meson.
		
	\end{itemize}
	\noindent
	\textcolor{black}{Note that in Ref.~\cite{QQQQ.ground:cbcb;ccbb.2019.1} the multiquark components in the ground states of such tetraquarks were studied in detail emphasizing the role played by the meson--meson and diquark--antidiquark configurations in the formation of the tetraquark states. The search for the resonances was performed with the real scaling method. In Ref.~\cite{Richard2017} the importance of studying such tetraquarks was claimed.}
	\\
	\hfill
	\\
	\noindent
	\textcolor{black}{The masses of the ground and excited states of tetraquarks presented in Ref.~\cite{qqQQ.ground:sbsb;ssbb;qqbb.QQQQ.ground;exc:cccc;cbcb;bbbb;ccbb.2020.1} are systematically lower than our predictions by about $100-400$ MeV. This is the result of different treatment of the relativistic effects and diquark structure (see the discussion above).}
	
	\item For the $\rm bb\overline b\overline c$, $\rm cb\overline b\overline b$ tetraquarks.
	\begin{itemize}

		\item In Ref.~\cite{qqqq;qqqQ;qqQQ;qQQQ;QQQQ.ground:all.1992.1} there is an indication of possible weak binding which is slightly stronger than for the $\rm cc\overline c\overline b$ tetraquark for the ground $J^{P}=1^{+}$ state (see Ref.~\cite{qqqq;qqqQ;qqQQ;qQQQ;QQQQ.ground:all.1992.1}, Fig.~8). The authors of Refs.~\cite{QQQQ.ground:all.2018.2, QQQQ.ground:all.2021.3} conclude that the lowest $J^{P}=1^{+}$ ground state can be narrow. Note that in Ref.~\cite{QQQQ.ground:all.2018.2} the lowest $J^{P}=1^{+}$ ground state is in the ($\overline 3 \times 3$)-$\rm A\overline S$ color--spin configuration. In Ref.~\cite{QQQQ.ground:all.2021.3} the corresponding state is in the ($6 \times \overline 6$)-$\rm S\overline A$ color--spin configuration. Despite this state lying above the fall-apart strong decay $\rm \eta_{b} \overline B_{c}$ channel such decay is forbidden due to the conservation of the angular momentum and parity. In addition, the authors of Ref.~\cite{QQQQ.ground:all.2021.4} predict a number of narrow resonances with fall-apart strong decay widths lying in the range of $1-30$ MeV (see Ref.~\cite{QQQQ.ground:all.2021.4}, Table~$\RNumb{16}$). Moreover, in Refs.~\cite{QQQQ.ground:all.2020.2, QQQQ.ground:all.exc:ccbb.2021.1, QQQQ.ground:cccb;ccbb;bbbc.2022.2} the tetraquark is predicted to be compact. 
		
		\item Still the authors of Refs.~\cite{QQQQ.ground:all.2019.3, QQQQ.ground;exc:cccc;bbbb;ccbb.2019.4, QQQQ.ground:all.2020.2, QQQQ.ground:cccb;ccbb;bbbc.2022.2, QQQQ.ground:all.2023.1} come to the conclusion that narrow bound ground states are not favored since all of the masses lie significantly above the lowest fall-apart strong decay thresholds. Yet there is a possibility for the resonances. Once again, in Ref.~\cite{QQQQ.ground:all.2020.2} the $J^{P}=0,1^{+}$ ground states are almost degenerate and the $J^{P}=2^{+}$ ground state lies $\approx$100 MeV above them. Note, that in Ref.~\cite{QQQQ.ground:all.2023.1} the ground state is dominated by the ($6 \times \overline 6$)-$\rm S\overline A$ color--spin configuration and considering only the hyperfine potential one can expect to have a narrow $J^{P}=1^{+}$ state which, however, still lies above the meson--meson threshold.
		
	\end{itemize}
	
\end{itemize}
Comparing the results discussed above with our calculations we come to the following conclusions.
\begin{itemize}
	
	\item Our results for the ground states of all compositions of tetraquarks agree with the arguments that these states lie highly above the meson--meson fall-apart thresholds (about $200-450$ MeV, see Tables~\ref{Tab:Comp.cccb.AA.1S}-\ref{Tab:Comp.bbbc.SA.1S}) and thus they fall-apart rapidly. 
	
	\item On the other hand
	\begin{itemize}
	
		\item For the $\rm cc\overline c\overline b$, $\rm bc\overline c\overline c$ and $\rm cc\overline b\overline b$, $\rm bb\overline c\overline c$ tetraquarks we find a few orbitally excited states (1P and 1D, to be precise) lying just above the thresholds, and one state (1D $J^{P}=4^{+}$) even below such a threshold (see Table~\ref{Tab:ThrProb} and Exprs.~\eqref{States:cccb},~\eqref{States:ccbb}), indicating a possibility of a long-living state. However, all these thresholds contain masses of the orbital excitations of the $B_{c}^{\pm}$-meson which were not observed yet but were theoretically calculated in Ref.~\cite{Regge2011meson}.
	
		\item For the $\rm bb\overline b\overline c$, $\rm cb\overline b\overline b$ tetraquarks we again find a few states right above the meson--meson fall-apart thresholds but no states below such thresholds (see Table~\ref{Tab:ThrProb}).
		
	\end{itemize}
		
\end{itemize}	
\noindent
It is worth noting, that the authors of Refs.~\cite{qqqq;qqqQ;qqQQ;qQQQ;QQQQ.ground:all.1992.1, tetrons2018, QQQQ.ground;exc:cccc;bbbb;ccbb.2019.4} did not predict any stable bound states for the fully charmed $\rm cc\overline c\overline c$ tetraquarks. However, a few states of this composition were already observed (see Table~\ref{Tab:ExpSym} and related discussion in Sec.~\ref{Sec:Thr}).

\section{Conclusion\label{Sec:Con}}

\par
We calculated masses of the ground (1S) and excited (1P, 2S, 1D, 2P, 3S) states of the asymmetric fully heavy tetraquarks on the basis of the relativistic quark model. The following compositions were studied: triple charmed and bottom ($\rm cc\overline c\overline b$, $\rm bc\overline c\overline c$), double charmed and double bottom ($\rm cc \overline b\overline b$, $\rm bb\overline c\overline c$) and triple bottom and charmed ($\rm bb\overline b\overline c$, $\rm cb\overline b\overline b$) tetraquarks.
\par
Our model is discussed in Secs.~\ref{Sec:Model},~\ref{Sec:Math}. The important feature of our calculations is the consistent account of the relativistic effects and the finite size of the diquark which leads to the weakening of the one-gluon exchange potential due to the form factors of the diquark-–gluon interaction.
\par
Additionally, a significant mixing of states with the same total momentum and parity $J^{P}$ but different full tetraquark spins $S$ occurs for the orbital excitations of the asymmetric compositions. It originates both from the spin--orbit and tensor terms of the quasipotential. In symmetrical compositions the mixing occurs from the tensor term only but just for a few orbitally excited states. Moreover, its magnitude is insignificant and thus this mixing can be completely neglected.
\par
A detailed analysis of the calculated mass spectra is presented in Sec.~\ref{Sec:Thr}. We compared calculated tetraquark masses with the thresholds of the fall-apart strong decays into the meson pairs. It is shown that most of the tetraquark states lie significantly above the threshold of fall-apart strong decay into a meson pair. However, a few tetraquark states are found to lie slightly above or even below such thresholds. As a result, these states are most likely to be observed as narrow states. An argument is given, as to why the excited states in general could be narrow despite the large phase space.
\par
We also present a comprehensive comparison of our results with the predictions of other theoretical calculations in Sec.~\ref{Sec:Theor}. We find that these results do not agree with each other in general. Nevertheless, almost all calculations indicate that the ground states of the asymmetric fully heavy tetraquarks lie significantly above the corresponding meson--meson thresholds.
\par
In conclusion, we note that experimental searches for the fully heavy tetraquarks are currently ongoing and should be continued. Therefore, it can be expected that new experimental candidates will appear in the near future.

\begin{acknowledgments}
The authors are grateful to D. Ebert and A.V. Berezhnoy for useful discussions. The work of Elena M. Savchenko was supported in part by the Foundation for the Advancement of Theoretical Physics and Mathematics ``BASIS'' grant number 22-2-10-3-1.
\end{acknowledgments}

\bibliography{Paper2024_revised.bib}

\end{document}